\documentclass[a4paper,fleqn]{cas-sc}

\usepackage[authoryear]{natbib}
\IfFileExists{stix.sty}{}{\usepackage{lmodern}}
\usepackage[safe]{tipa}
\usepackage{threeparttable}
\usepackage{arydshln}   %
\usepackage{xurl}      %

\newcommand{\Fone}[1]{F1@#1}
\newcommand{\ipa}[1]{\textipa{#1}}

\begin{document}
\let\WriteBookmarks\relax
\def\floatpagepagefraction{1}
\def\textpagefraction{.001}

\shorttitle{BFA: multilingual neural forced alignment}
\shortauthors{A. Rehman et al.}

\title [mode = title]{BFA: Multilingual Neural Forced Alignment with Boundary-Trained Phone Posteriors}

\author[1]{Abdul Rehman}[orcid=0000-0003-2345-2256]
\cormark[1]
\ead{_abdul@outlook.com} 
\credit{Conceptualization, Methodology, Software, Investigation, Formal analysis, Data curation, Writing -- original draft}

\author[1]{Jingyao Cai}[orcid=0009-0005-3655-4915]
\credit{Formal analysis, Data curation,}

\author[1]{Jian Jun Zhang}[orcid=0000-0002-7069-5771]
\credit{Supervision, Funding acquisition, Writing -- review \& editing}

\author[1]{Xiaosong Yang}[orcid=0000-0003-3815-0584]
\credit{Validation, Resources, Supervision, Funding acquisition, Writing -- review \& editing}

\affiliation[1]{organization={Bournemouth University},
            addressline={Fern Barrow},
            city={Poole},
            postcode={BH12 5BB},
            country={United Kingdom}}

\cortext[1]{Corresponding author}

\begin{abstract}
Phone-level forced alignment underlies phonetic measurement, corpus annotation and speech synthesis. Established aligners need a pronunciation dictionary and an acoustic model for each language, and neural aligners built on connectionist temporal classification (CTC) recognisers mainly serve word-level timing. This paper asks whether a neural aligner can place phone boundaries in many languages, including ones it was not trained on, without a dictionary or boundary labels. In the Bournemouth Forced Aligner (BFA), hat-shaped soft targets make adjacent phone posteriors cross at their shared boundary. The targets come from the model's own segmental Viterbi alignment, refined from a flat start, so only transcripts are needed; the same search, with a soft duration prior and an acoustic-change onset term, decodes at inference. A grapheme-to-phoneme front end replaces the dictionary, with one compact inventory per writing system. The convolutional acoustic model hears only a few tens of milliseconds of audio per frame, so it learns how phones sound rather than which language they belong to. We evaluate against hand-corrected boundaries on read and conversational English. Models that never heard English align English nearly as well as models that did, suggesting that a language can be aligned once its phones have been heard in others. The boundary objective and the segmental search together give the sharp boundaries that phonetic work requires. We also measure alignment when the transcript is wrong or the audio contains untranscribed speech, and find the damage stays close to the error.
\end{abstract}

\begin{highlights}
\item Multilingual phone alignment without a pronunciation dictionary or boundary labels
\item Relative to CTC alignment, BFA gives sharper boundaries for phonetic work
\item A model that never heard English aligns English nearly as well as one that did
\item Alignment under transcript errors and audio corruption is tested against hand labels
\end{highlights}


\maketitle

\section{Introduction}\label{sec:intro}

Phone-level timestamps are the input to a wide range of speech work: duration and prosody targets for speech synthesis, phonetic and sociolinguistic measurement, the time-locking of neural responses to speech \citep{agmon2025challenges}, pronunciation assessment, corpus annotation and lip-sync. The standard tool is a GMM--HMM forced aligner, such as the HTK-based aligners \citep{young2006htk,gorman2011prosodylab,yuan2008p2fa} and, most widely today, the Montreal Forced Aligner \citep[MFA;][]{mcauliffe2017mfa}. These aligners are accurate on the languages and varieties they were built for \citep{mackenzie2020british,williams2024l2}, but each language needs (i)~a pronunciation dictionary that covers the transcript's vocabulary and (ii)~an acoustic model trained on that language. For most of the world's languages neither exists \citep{dicanio2013}. Mapping a new language's phones onto an existing aligner's models works around this. Where it has been scored on phone boundaries, it works well for a target language close to the source \citep{young2023forced} and less well for a distant one \citep{dicanio2013}, and many evaluations score only word edges or compare with another aligner's output rather than with hand labels \citep{strunk2014webmaus,leinonen2021grapheme,wu2025crossphon} (Section~\ref{sec:related}).

Neural alternatives mostly reuse recognisers trained with connectionist temporal classification \citep[CTC;][]{graves2006ctc,kurzinger2020ctcseg,rastorgueva2023nfa,pratap2024mms}. Aligners of this kind, such as the MMS aligner \citep{pratap2024mms} and WhisperX \citep{bain2023whisperx}, mainly serve word- or character-level timing for subtitles, long-form transcription and the preparation of speech-recognition data. There a word edge within a few tens of milliseconds is good enough, and they cover far more languages than any HMM aligner. Phonetic work asks for phone boundaries within 10--25\,ms, about the range within which human labellers agree with each other \citep{young2023forced}, and analyses of neural responses to speech degrade when phone onsets are off by more \citep{agmon2025challenges}; such boundaries are harder to obtain from a CTC model, because CTC posteriors tend to be \emph{peaky} \citep{zeyer2021peaky,huang2024less}: each phone is marked by a spike of one or two frames, and a blank class covers the frames in between. The CTC objective does not say where in a blank run one phone ends and the next begins, so the decoder has to estimate the boundary afterwards, for example from the first frame of the phone on the Viterbi path. In our experiments this gives boundaries that are not sharp enough for phonetic use (Table~\ref{tab:main} and Section~\ref{sec:res-trail}).

This paper asks whether a neural aligner can provide phone boundaries for many languages, including ones it was not trained on, without a dictionary or boundary labels for any of them. Two things make this possible. The boundary itself can be trained into the posterior without ever showing the model a boundary label, with a training objective in which adjacent phones cross at their shared boundary and a decoder that is the same in training and at inference. And an acoustic model that sees only a few tens of milliseconds of audio learns how phones sound rather than which language they belong to, so that, with a grapheme-to-phoneme front end in place of the dictionary, one model can serve every language of a writing system. We test the second claim directly, by leaving English out of training and aligning English. For the other script groups we check the models against a human-verified reference that BFA helped to seed, and so is partial to it (Section~\ref{sec:res-wild}).

An aligner for languages without curated resources will often be given imperfect input, so we also ask \emph{what happens when the transcript is wrong?} In use the text often comes from automatic speech recognition, which tends to drop disfluencies and repairs \citep{agmon2025challenges}, from a G2P or dictionary pronunciation of reduced speech, or from a hurried transcriber. A forced aligner must place every phone it is given, so each error displaces real boundaries. We measure this against hand labels, with phone-level and word-level errors at controlled rates, and ask the converse question, \emph{what happens when the audio contains something the transcript does not?}, by inserting and replacing stretches of audio under an unchanged transcript.

Our contributions are:
\begin{enumerate}
\item \textbf{A blank-free boundary objective for a neural aligner} (Section~\ref{sec:model}). Hat-shaped soft targets make two neighbouring phone posteriors cross at 0.5 at their shared boundary. The targets are built from the model's own segmental Viterbi alignment, by hard EM from a flat start, so that only transcripts are needed, as for HMM and CTC aligners.

\item \textbf{A decoder shared by training and inference} (Section~\ref{sec:decoding}): a segmental Viterbi search with a soft duration prior, scaled-likelihood emissions and an acoustic-change onset term inside the global search. We show that the onset term helps on read speech and can hurt on conversational speech.

\item \textbf{A dictionary-free phone inventory organised by writing system} (Section~\ref{sec:inventory}), with one compact token list and one output head per script group, and evidence that the acoustic model does not depend on the language it aligns: a model trained without English aligns English nearly as well as one trained with it (Section~\ref{sec:res-unseen}).

\item \textbf{A public release} of the aligner, the models and the evaluation harness, with the configuration of every reported run. The development history is reported here and made public, so that the results can be reproduced and the models can be improved without retraining from scratch (Section~\ref{sec:res-trail}).
\end{enumerate}

\subsection{The method in brief}\label{sec:brief}

The input is a recording and its transcript, for example the word \emph{cat}. A grapheme-to-phoneme model turns the text into its phone sequence, here \ipa{/k/}, \ipa{/\ae/}, \ipa{/t/}, and the task is to cut the time axis into consecutive pieces, one per phone, in that order. A small convolutional network reads the audio in frames of 5\,ms, each with a few tens of milliseconds of context, and outputs a probability for every phone at every frame. Dynamic programming then finds, among all ways of assigning the frames to the phones in order, the one with the highest total score. The score also penalises implausibly short phones and rewards starting a phone where the spectrum changes abruptly (the onset bonus $\beta$).

Training has no timing labels, only transcripts. As in HMM training, we start from an even split of each recording among its phones and alternate two steps: the network is trained on the current segmentation, and the improved network segments the recordings again with the same dynamic programme. What is specific to BFA is the training target. Instead of switching from one phone to the next at the boundary, the target fades linearly between them over a few frames, so that the two probabilities cross at 50\% at the boundary. A trained network therefore marks each boundary by a crossing of two posterior curves (Fig.~\ref{fig:hat}), and that is where the decoder places it. Nothing in this procedure relies on a pronunciation dictionary or on a particular language.

The acoustic trunk is a contextless CNN of the CUPE family \citep{rehman2025cupe}, which we do not re-evaluate as a phoneme recogniser (using scores such as Phone Error Rate). The objective, the decoder, the inventory and the evaluation are new. The previous releases of BFA (v1.X) used a CUPE-2i trunk, which is now superseded by the CUPE-3-en trunk of v2.X. The major development steps that increased the boundary accuracy were replacing the CTC objective with the hat objective, and replacing the CTC decoder with the segmental Viterbi search. This paper mainly evaluates BFA v2.1, but the v1.1.5 library and models are also included in the tables for comparison. The other differences are new language groups, made possible by a neural G2P, and smaller models, from removing the transformer layers of CUPE-2i entirely: they helped phoneme recognition but did not sharpen boundaries.

For comparison with other aligners, we treat TIMIT \citep{garofolo1993timit} as the most reliable source of hand-labelled boundaries, and Buckeye \citep{pitt2005buckeye} as a more realistic test of conversational speech. We also check the script-group models on WildPhonemes \citep{ar2026_wildphonemes}, a new small set in 14 languages. The field has no common benchmark for forced alignment, and published evaluations differ in corpora, tolerances and metrics \citep{deng2025speech}. This paper is not a benchmark of aligners either, and the other aligners are not ranked: mismatches in their conventions, dictionaries, acoustic models and training data make a fair comparison difficult. For example, the junction between two identical vowels (\ipa{/i/}--\ipa{/i/}), or whether a phone covers breath sounds or trailing echoes, is usually a matter of convention, not accuracy.

\section{Related work}\label{sec:related}

\paragraph{HMM forced alignment.} HTK-style aligners and their descendants, among them Prosodylab-Aligner \citep{gorman2011prosodylab}, P2FA \citep{yuan2008p2fa}, WebMAUS \citep{kisler2017maus}, SPPAS \citep{bigi2012sppas} and MFA \citep{mcauliffe2017mfa}, train GMM--HMMs by embedded re-estimation from a flat start and align with Viterbi decoding under a pronunciation lexicon. MFA, built on Kaldi \citep{povey2011kaldi}, refines a monophone model into context-dependent (triphone) models with speaker adaptation; it is the de facto standard and publishes benchmarks on TIMIT and Buckeye \citep{mfamodels}. Systems trained on TIMIT's own boundaries reach high agreement with its labels \citep{hosom2009,stolcke2014}, but need hand-segmented data in the target language. Our training loop borrows the flat-start, hard-EM bootstrap of this line \citep{juang1990segmental}; the difference is that the emission model is a discriminative CNN trained towards boundary-shaped targets rather than a generative GMM.

\paragraph{Cross-language forced alignment.} A language without an acoustic model of its own is usually aligned with another language's models, after its phones or graphemes are mapped onto theirs. This has been evaluated with English models on endangered and under-resourced languages \citep{dicanio2013} and on Swedish \citep{young2023forced}, with WebMAUS's language-independent mode on language documentation corpora \citep{strunk2014webmaus}, and with Finnish models on other Uralic languages \citep{leinonen2021grapheme,leinonen2022semiautomatic}. Accuracy depends on the pair of languages: tonality and the phone inventory matter in choosing the source model \citep{wu2024cross}, and phones without a counterpart in the source align worse \citep{chodroff2025comparing}. A large and varied English model can match language-specific models trained on small amounts of target-language data \citep{chodroff2025comparing}. Each of these set-ups rests on a phone mapping made by hand, which CrossPhon derives instead from phonological features \citep{wu2025crossphon}. BFA differs in two respects: its source is not one language but every language of a script group, and the mapping is made once, by the back-off rules of the group's inventory (Section~\ref{sec:inventory}), rather than for each target language.

\paragraph{Hybrid and segmental models.} Dividing network posteriors by class priors to obtain scaled likelihoods for HMM decoding is classical \citep{bourlard1994}, and we use it for the same reason. Explicit duration modelling with segment models and hidden semi-Markov models has a long history \citep{ostendorf1996segment,yu2010hsmm}; our decoder is a light-weight instance with a soft duration floor (Section~\ref{sec:topology}).

\paragraph{CTC-based aligners.} CTC segmentation \citep{kurzinger2020ctcseg}, the NeMo Forced Aligner \citep{rastorgueva2023nfa} and the MMS aligner \citep{pratap2024mms} align with CTC models, usually at character or subword level. Phoneme-level CTC models, such as wav2vec~2.0 \citep{baevski2020wav2vec2} fine-tuned on multilingual phonemes \citep{xu2022zeroshot}, one of our baselines (Section~\ref{sec:systems}), and Allosaurus \citep{li2020allosaurus}, provide multilingual phone posteriors. WhisperX \citep{bain2023whisperx} times the words of a Whisper transcript with a CTC model fine-tuned from wav2vec~2.0. These systems are built for word or subword timing over many languages, which they deliver at scale; \citet{rousso2024tradition} compare them with MFA on word boundaries, and in comparisons on word boundaries CTC segmentation has generally been less precise than HMM alignment \citep{leinonen2022semiautomatic,deng2025speech}. Phone boundaries ask more of the posteriors, because CTC's peaky output \citep{zeyer2021peaky, huang2024less} leaves the position of a boundary inside a blank run to the decoder.

\paragraph{Neural phone aligners.} Charsiu \citep{zhu2022charsiu} offers two aligners built on wav2vec~2.0. Its frame classifier, W2V2-FC, is trained on alignments produced by MFA, and the Mason-Alberta Phonetic Segmenter \citep[MAPS;][]{kelley2024maps} is trained on hand-labelled boundaries; both inherit their boundaries from outside the model. Charsiu's W2V2-FS instead learns a monotonic attention between audio frames and the transcript's phones from transcripts alone, with a forward-sum loss \citep{badlani2022onetts}; it is the closest neural aligner to ours in supervision, and one of our baselines (Section~\ref{sec:systems}).

\paragraph{Alignment learned from transcripts alone.} Other neural models also learn alignments without boundary labels. Text-to-speech models learn a monotonic alignment between text and speech frames together with synthesis, by monotonic alignment search in Glow-TTS \citep{kim2020glowtts} or by a forward-sum objective \citep{badlani2022onetts}, and NeuFA \citep{li2022neufa} derives a forced aligner from the attention between a synthesis and a recognition task. Unsupervised phone segmentation \citep[e.g.][]{kreuk2020segmentation} finds boundaries from changes in learned representations without any text, and so does not assign them to the phones of a transcript. BFA shares the transcript-only supervision of the first group and the acoustic-change cue of the second (Section~\ref{sec:onset}). It differs in that the boundary is trained into frame-level phone posteriors on a 5\,ms grid by the same segmental search that decodes them, and in that the model bootstraps from its own alignments, as an HMM aligner does.

\paragraph{Multilingual G2P and inventories.} We phonemise with a ByT5 grapheme-to-phoneme (G2P) model covering about 100 languages \citep{zhu2022byt5g2p}, post-processed into a curated inventory (Section~\ref{sec:inventory}). Shared multilingual phone sets, such as Allosaurus' allophone layer \citep{li2020allosaurus}, map a universal inventory to per-language phonemes. We go the other way: a small inventory per script, plus a universal broad-class head.

\section{Model and training objective}\label{sec:model}

\subsection{Acoustic encoder}\label{sec:encoder}

The trunk follows CUPE \citep{rehman2025cupe}; we state only what differs. The model reads raw 16\,kHz audio in 120\,ms windows with an 80\,ms stride. A strided convolutional front end (kernels 15/9/7, strides $4\cdot4\cdot5$) gives one frame per 80 samples (5\,ms), two residual context convolutions (kernels 5 and 3) bring the receptive field to 38.9\,ms, and four per-frame MLP blocks with no time mixing follow. Window outputs are overlap-added with a cosine taper. Every kernel is odd with symmetric padding, so frame $t$ is centred on sample $80t$, at exactly $5t$\,ms, and timestamps need no offset correction.

A receptive field of 39\,ms is far too short to learn phonotactics or a lexicon. This is what lets one trunk transfer across languages \citep{rehman2025cupe}, and it makes the output at a frame independent of how much audio surrounds it. For the same reason the trunk omits the transformer layers of CUPE: self-attention lets every frame draw on the whole window, which helps phoneme recognition, whereas a boundary should be decided by the audio at the boundary. A probe of the layers of a CTC model of this family, and an attention layer added to the trunk, both bear this out (Section~\ref{sec:res-trail}).

\paragraph{Output heads.} Two model families share the 12.8\,M-parameter trunk and differ in their phone inventory (Section~\ref{sec:inventory}). The \emph{ph66} models have one 66-phoneme head, shared by all languages. The \emph{std} models have a \emph{local} phoneme head over the tokens of one script group. Both families also have a small broad-class head (15 or 16 classes such as vowels, stops and nasals), trained as an auxiliary task, in order to help the trunk learn phonetic features that transfer across languages. The broad-class head is not used at inference, and the std models have one further auxiliary head that this paper does not use (Section~\ref{sec:limitations}).

\subsection{The boundary as a posterior crossing}\label{sec:hat}

Let a clip have frames $t = 0,\dots,T-1$ and a known phone sequence $c_1,\dots,c_N$, and let $p(c\mid x_t)$ be the network's posterior for class $c$ at frame $t$, where $x_t$ is the audio in the frame's receptive field. Suppose an alignment gives onset frames $o_1 < \dots < o_N$, with the boundary between phones $i-1$ and $i$ at $b_i = o_i - \tfrac12$, between frames $o_i-1$ and $o_i$. For a frame $t$ in phone $i$, define its distances to the two boundaries and a weight
\begin{equation}\label{eq:hat}
d^{L}_t = t - b_i,\qquad d^{R}_t = b_{i+1} - t,\qquad
w_t = \frac12 + \frac12 \min\!\left(\frac{\min(d^{L}_t, d^{R}_t)}{r},\, 1\right).
\end{equation}
The target distribution puts $w_t$ on $c_i$ and $1-w_t$ on the nearer neighbour. Thus $w_t = 1$ in the interior of the phone and falls linearly to 0.5 at each boundary, where the neighbour rises through 0.5. A phone shorter than $2r$ becomes a triangle that peaks below~1. The ramp $r$ is 1 frame (5\,ms) by default for the std models, except mbfa-latin-benchmark ($r = 3$) and mbfa-latin-noEnglish ($r = 4$).

\begin{figure}[pos=!htb]
  \centering
  \includegraphics[width=\textwidth]{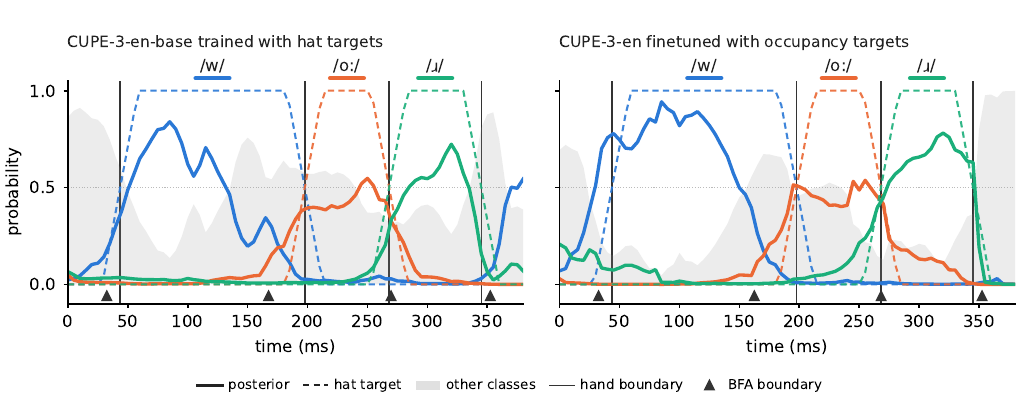}
  \caption{Hat targets and posteriors over \ipa{/w O: \*r/} of \emph{worn} in a TIMIT test clip (FAKS0/SI1573), at the decoder defaults. Left: CUPE-3-en-base, trained on the hat ($r = 3$). Right: CUPE-3-en, the same network family trained on occupancy targets instead (Section~\ref{sec:selfalign}), for comparison. Dashed: the hat target ($r = 3$ frames) built from the hand-labelled boundaries (vertical lines). Solid: the model's posterior for the same three classes. Grey: the mass on every other class. Triangles: BFA's boundaries with the sub-frame read-out (Supplementary Section~S7). In both, adjacent posteriors cross near the boundaries and none saturates. The \ipa{/O:/}--\ipa{/\*r/} boundary is read within 2\,ms of the hand label. The \ipa{/w/}--\ipa{/O:/} glide is gradual, and there the posteriors cross, and BFA places the boundary, 30--35\,ms before the hand label. The clip was picked by a fixed rule, not by eye: the first test clip, by ID, with three consecutive hand-labelled vowels, fricatives, nasals or approximants of at least 60\,ms each that BFA aligns to the same three phones.}
  \label{fig:hat}
\end{figure}

Two consequences matter for alignment (Fig.~\ref{fig:hat}):
\begin{itemize}
\item \emph{In the target, a boundary is where two adjacent posteriors cross}, and nowhere else. A trained model reproduces this approximately, so a boundary can be read between frames, as the zero of $\log p(c_i\mid x_t) - \log p(c_{i-1}\mid x_t)$ (Supplementary Section~S7).
\item \emph{Posteriors do not saturate.} The target has entropy by design, so the cross-entropy cannot reach zero (only $\mathrm{CE} - H(\text{target})$ can). A correctly aligned phone often has a peak posterior below 0.5, which changes how confidence must be read (Section~\ref{sec:conf}). The hat is in this sense a structured form of label smoothing \citep{szegedy2016rethinking}, applied only where the labels are uncertain: near the boundaries.
\end{itemize}

\subsection{Self-alignment without boundary labels}\label{sec:selfalign}

The onsets $o_i$ are never taken from annotation. At every training step:
\begin{enumerate}
\item \textbf{Align} (no gradient). The segmental Viterbi search of Section~\ref{sec:decoding} is run on the model's current posteriors and assigns every frame to one phone of the known sequence. The search includes the acoustic-change onset term of Section~\ref{sec:onset}, so the targets are placed with acoustic evidence that the posteriors do not contain; Supplementary Section~S8 tests training without it.
\item \textbf{Build targets} as in Section~\ref{sec:hat} and minimise the cross-entropy on the phoneme and broad-class heads,
\begin{equation}
\mathcal{L} = \mathrm{CE}_{\text{ph}} + 0.3\,\mathrm{CE}_{\text{group}}.
\end{equation}
The std models carry one further auxiliary head that this paper does not use (Section~\ref{sec:limitations}).
\end{enumerate}
This is hard EM from a flat start, the bootstrap of HMM aligners \citep{juang1990segmental,young2006htk}. On a cold start a diagonal prior pulls the path towards an even split by phone count, and its weight is annealed to zero over the first epochs. Training and decoding share the search, not necessarily its settings: the training runs of the ph66 models used $D = 3$ or 6 and $K = 4$ or 10, whereas every model is decoded at the defaults of Section~\ref{sec:sweep}.

\paragraph{Where the alignment's uncertainty goes.} A forward--backward pass of the model's posteriors over the same topology gives each frame's \emph{occupancy}: the probability that it lies in each phone of the sequence, summed over every segmentation the search allows. The models use it in one of two ways. The std models use it only to move the boundaries: each is placed at its expected, fractional position instead of at the Viterbi onset, and the hat of Eq.~\ref{eq:hat} is built around that position. The target's shape is still set by $r$ alone, so it does not change with the model's confidence. At $r = 1$, the std default, the target is close to a hard label that reaches 1 one frame from the boundary, and the weights of the two frames beside the boundary record where between them it lies, which a hard label on the frame grid cannot. A matched comparison showed no gain in accuracy from the fractional positions on the TIMIT probe. CUPE-3-mls7 and CUPE-3-en instead replaced the hat with the occupancy itself, gathered onto each phone's class. That target is only as soft as the model is uncertain, so a more confident model is given sharper targets, which make it more confident still. Over long runs, TIMIT boundary \Fone{25} drifted down while the training loss still fell (Supplementary Section~S9). CUPE-3-en-base and every std model therefore use the hat, whose shape the model cannot change.

\paragraph{Noise.} Gaussian noise of standard deviation $\sigma$ is added to the RMS-normalised waveforms during training, to make the models less sensitive to recording conditions. The level is highest ($\sigma = 0.05$) for the two models trained on the full Latin-script data, mbfa-latin and mbfa-latin-noEnglish, which cover the most languages (Supplementary Section~S9).

\subsection{Training data and released models}\label{sec:data}

The ph66 models are trained on read audiobooks: LibriSpeech \citep{panayotov2015librispeech} and 100\,h from each of the seven non-English languages of Multilingual LibriSpeech \citep[MLS;][]{pratap2020mls} (German, Dutch, French, Spanish, Italian, Portuguese and Polish). The std models are trained on FLEURS \citep{conneau2023fleurs}, read speech in 102 languages with about 12\,h per language, whose speakers and recording conditions are more varied than an audiobook's. Segments whose transcript seems not to match the audio are filtered out, which leaves 2.7--9.7\,h per language (Supplementary Sections~S9 and~S10). Every transcript is phonemised by the front end that is used at inference (eSpeak NG for ph66, the ByT5 G2P for std), so the models learn from G2P pronunciations, not hand-checked ones.

Table~\ref{tab:models} lists the models. CUPE-3-en-base is trained on LibriSpeech only, to match the training data of MFA's LibriSpeech-trained ARPAbet model. CUPE-3-mls7 was trained from random weights on the seven MLS languages and never heard English; CUPE-3-en continues it on LibriSpeech, so it is English-dominant rather than English-only. Both were trained with occupancy targets instead of the hat (Section~\ref{sec:selfalign}). mbfa-latin-noEnglish was trained from random weights on the Latin-script data with English removed. The non-Latin std models were initialised from the from-scratch Latin run, keeping the trunk and the broad-class head; their boundaries are checked on 14 languages of Common Voice against a human-verified reference that is partial to BFA (Section~\ref{sec:res-wild}). The optimiser settings, checkpoint selection and the training lineage of every model are in Supplementary Section~S9.

\begin{table}[t]
\caption{Released models. ph66 models are trained on LibriSpeech \citep{panayotov2015librispeech} and MLS \citep{pratap2020mls}; CUPE-3-en-base and CUPE-3-mls7 start from random weights, and CUPE-3-en continues CUPE-3-mls7 (``+''); std models on FLEURS \citep{conneau2023fleurs}, with hours after filtering.}\label{tab:models}
\small
\begin{tabular*}{\tblwidth}{@{\extracolsep{\fill}}llll@{}}
\toprule
Model & Inventory (tokens) & Training data & Params \\
\midrule
CUPE-3-en-base & ph66 & LibriSpeech, 100\,h then $\sim$500\,h & 13.54\,M \\
CUPE-3-mls7 & ph66 & MLS de es fr it nl pl pt, 100\,h each (no English) & 13.54\,M \\
CUPE-3-en & ph66 & + LibriSpeech 360\,h and $\sim$500\,h & 13.54\,M \\
\midrule
mbfa-latin & std/latin (208) & 36 Latin-script languages, 273.5\,h & 14.88\,M \\
mbfa-latin-benchmark & std/latin (208) & the same, a 62\% sample, $\sim$170\,h & 14.88\,M \\
mbfa-latin-noEnglish & std/latin (208) & 35 languages without English, 266.4\,h & 14.88\,M \\
mbfa-cyrillic & std/cyrillic (127) & 8 languages, $\sim$40\,h & 14.79\,M \\
mbfa-cjk & std/cjk (54) & Mandarin, Cantonese, $\sim$9\,h & 14.71\,M \\
mbfa-japanese & std/japanese (28) & Japanese, 7\,h & 14.69\,M \\
mbfa-vietnamese & std/vietnamese (45) & Vietnamese, 8.7\,h & 14.71\,M \\
mbfa-thai & std/thai\_khmer (38) & Thai, 7.8\,h & 14.70\,M \\
mbfa-abjad & std/abjad (90) & Arabic, Central Kurdish, Persian, Sindhi, Urdu, 38.2\,h & 14.75\,M \\
mbfa-brahmic & std/brahmic (96) & Hindi, Odia, Tamil, 16.8\,h & 14.76\,M \\
mbfa-other-alphabetic & std/other\_alphabetic (104) & Amharic, Greek, Armenian, Georgian, Korean, 39.8\,h & 14.77\,M \\
\bottomrule
\end{tabular*}
\end{table}

\section{Decoding}\label{sec:decoding}

\subsection{Segmental search with a soft duration prior}\label{sec:topology}

Given per-frame emission scores $e_t(c)$ (Section~\ref{sec:emissions}) and an onset score $a_t$ (Section~\ref{sec:onset}), the decoder chooses onset frames $\mathbf{o} = (o_1,\dots,o_N)$ for the known phone sequence by
\begin{equation}\label{eq:search}
\hat{\mathbf{o}} = \arg\max_{\mathbf{o}} \sum_{i=1}^{N}\Big[\sum_{t=o_i}^{o_{i+1}-1} e_t(c_i) \;-\; \kappa\,\max(0,\,D+K-\ell_i) \;+\; \beta\,a_{o_i}\Big],
\qquad \ell_i = o_{i+1}-o_i \ge D,
\end{equation}
with $o_{N+1} = T$. Here $\ell_i$ is the duration of phone $i$ in frames, $D$ is a hard minimum duration, $K$ is the number of further frames over which a short phone is penalised, $\kappa$ is the penalty per missing frame, and $\beta$ is the weight of the onset term. A phone shorter than $D$ frames is infeasible, a phone of $D+K$ frames or more pays nothing, and in between the penalty falls linearly. This is a \emph{soft} duration prior, which a constant per-phone penalty cannot be: every path enters the same number of phones and would pay a constant equally. The search is a Viterbi pass over an HMM in which each phone is a left-to-right chain of $D+K$ sub-states. The first $D$ are mandatory, each of the next $K$ can be skipped at a cost $\kappa$, and only the last has a self-loop, so the search costs $O(TN(D+K))$. The defaults are $D = 6$ frames (30\,ms), $K = 10$ and $\kappa = 1$\,nat (Section~\ref{sec:sweep}).

Silence is handled by optional states. One is added at each end of the segment and may be skipped at no cost, because the phonemiser rarely writes leading silence and, without a blank class, nothing else would absorb it; interior silences are skippable and need one frame if visited. If a segment's text cannot fit its audio at the floor $D$, the floor is lowered to $\lfloor T/N \rfloor$ for that segment (with $K = 0$), so the segment is aligned with a weaker prior rather than dropped. The resulting segmentation is contiguous: every frame belongs to one phone or to a silence, so a phone's offset is its successor's onset.

The onsets $o_i$ lie on the 5\,ms frame grid, and every result in this paper uses them as they are. Interpolating the crossing of the two adjacent log-posteriors between frames gives a sub-frame estimate; on a 1,000-clip TIMIT subset with the onset term on it raises \Fone{10} by under one point and leaves \Fone{25} essentially unchanged (Supplementary Section~S7), and we use it only to draw Fig.~\ref{fig:hat}.

\subsection{Emissions: scaled likelihoods}\label{sec:emissions}

Viterbi decoding on raw posteriors rewards giving frames to whichever class the model favours everywhere. Following hybrid NN--HMM practice \citep{bourlard1994}, we score
\begin{equation}\label{eq:emission}
e_t(c) = \log p(c\mid x_t) - \log \bar p(c),\qquad
\bar p(c) = (1-\lambda)\,\frac{1}{|\mathcal{T}|}\sum_{t\in\mathcal{T}} p(c\mid x_t) + \frac{\lambda}{C},
\end{equation}
with $\mathcal{T}$ the valid frames of the batch, $C$ the number of classes and $\lambda = 0.1$. Dividing by the prior turns ``how likely is this phone here'' into ``how much more likely is it here than elsewhere'', which is what a boundary decision needs. Decoding raw log-posteriors instead costs under one point of F1 at both tolerances on a 1,000-clip TIMIT subset, for both model families.

\subsection{Acoustic-change onset term}\label{sec:onset}

Let $f_t = \lVert \mathbf{m}_t - \mathbf{m}_{t-1}\rVert_2$ be the change of an 80-band log-mel spectrum $\mathbf{m}_t$ (25\,ms window, 5\,ms hop, on the model's frame grid). The onset score is the local contrast
\begin{equation}\label{eq:onset}
a_t = \frac{f_t - \mathrm{MA}_{W}(f)_t}{\mathrm{SD}(f)},
\end{equation}
with $\mathrm{MA}_W$ a moving average over $W$ frames and $\mathrm{SD}(f)$ the standard deviation of $f$ over the segment. The search adds $\beta a_t$ nats to every transition that \emph{starts} a phone at frame $t$ (Eq.~\ref{eq:search}), so posterior evidence and acoustic change are traded off in one global search. Local contrast is used rather than raw spectral flux \citep{bello2005onset} because loud speech has high flux everywhere. Since $a_t$ is negative where the spectrum changes less than its surroundings, the term also discourages boundaries in steady stretches (e.g., vowel--vowel junctions).

\subsection{Choice of decoder settings}\label{sec:sweep}

The defaults ($D = 6$, $K = 10$, $\kappa = 1$, $W = 33$, $\beta = 15$ for every model) were chosen by a sweep over a seeded 1,000-clip TIMIT subset and ten Buckeye recordings, with one ph66 and one std model; the full grid and a reading of each setting are in Supplementary Section~S1. These subsets come from the corpora that Section~\ref{sec:results} scores, so the defaults are tuned on part of the test data, which is worth about one point of \Fone{25} or less. Read and conversational speech disagree on two settings: TIMIT, many of whose phones are short, prefers a shorter duration floor than Buckeye, and read speech favours a larger onset weight than conversational speech (Supplementary Sections~S1 and~S8).

\subsection{Confidence}\label{sec:conf}

A phone's confidence is its \emph{peak} posterior inside its own segment; the mean would penalise short phones, which under the hat target are nearly all ramp. Because the hat keeps posteriors below~1, confidence scales differ across models: the ph66 models are about twice as confident as the std models on TIMIT, so a threshold does not transfer from one family to the other. Section~\ref{sec:res-errors} tests what confidence predicts.

\section{A dictionary-free inventory by writing system}\label{sec:inventory}

\paragraph{From text to tokens.} Text is split into words, by a deterministic segmenter for scripts written without spaces, and each word is transcribed to IPA by the ByT5 G2P \citep{zhu2022byt5g2p}. The IPA is normalised into a curated inventory and then into the \emph{local token list of the language's script group}. A phoneme missing from a group's list backs off to its nearest neighbour, to two tokens (affricates and diphthongs: \ipa{/tS/} $\to$ \ipa{t S}, \ipa{/aI/} $\to$ \ipa{a I}), or as a last resort to an unknown token. No lexicon is consulted when aligning, so the user supplies none and no word is out of vocabulary, at the price of G2P errors (Section~\ref{sec:limitations}). This is the sense in which we call the aligner dictionary-free. The front ends carry lexical knowledge of their own: the ByT5 G2P was trained on pronunciation lexicons, and eSpeak NG (Section~\ref{sec:ph66}) is built from pronunciation rules and exception lists for each language.

\paragraph{Grouping by writing system.} A single universal inventory spends output classes on contrasts that a given language never makes, and a per-language inventory needs a model per language. Grouping by script is a practical middle ground: languages sharing a script tend to share much of a G2P's output inventory, a group's data can be pooled, and the head stays small (28--208 tokens, Table~\ref{tab:models}). All groups share the 15-class broad-class head (vowels by height, stops and fricatives by voicing, nasals, laterals, rhotics, glides, and so on), which is kept when a new group's model is initialised from an existing one. The training data cover 62 languages in nine groups. A further 20 languages can be phonemised into a group's tokens without having been in its training data; we make no accuracy claim for them. The groups and their languages are listed in Supplementary Section~S10.

\paragraph{The ph66 inventory.}\label{sec:ph66} The second inventory phonemises with eSpeak NG \citep{espeakng} and reduces its output to a fixed set of 66 phonemes shared by all supported languages. It is decoded by the ph66 models, which are the more accurate of the two families on English (Section~\ref{sec:res-unseen}).

\section{Experimental setup}\label{sec:setup}

\subsection{Corpora}

\begin{table}[t]
\caption{Corpora: TIMIT \citep{garofolo1993timit} and Buckeye \citep{pitt2005buckeye,pitt2007buckeye}, on which boundary accuracy is evaluated, and WildPhonemes \citep{ar2026_wildphonemes}, used only as a check of the non-English models because BFA helped to seed its reference (Section~\ref{sec:res-wild}). All three are scored in full.}\label{tab:corpora}
\small
\begin{tabular*}{\tblwidth}{@{}p{0.13\textwidth}p{0.23\textwidth}p{0.27\textwidth}p{0.31\textwidth}@{}}
\toprule
 & TIMIT & Buckeye & WildPhonemes (check only) \\
\midrule
Language & English & English & 14 languages, at least one per script group \\
Speech & read, studio & conversational interviews & read prompts, crowd-sourced (Common Voice 22.0) \\
Scored & all 6,300 clips (TRAIN + TEST) & all 255 recordings, 40 speakers, 33,676 segments of $\sim$2.3\,s & all 1,203 clips (125.6\,min), 1,166 speakers \\
Reference boundaries & 223,864 phone onsets & 840,116 phone onsets & 70,356 phone onsets, 29,878 word edges \\
Labelled & automatic alignment, corrected by hand & HMM aligner, then hand-corrected & word edges seeded at the midpoint of BFA and MFA, phones force-aligned inside each word by BFA; word edges and phone boundaries then hand-corrected \\
\bottomrule
\end{tabular*}
\end{table}

None of the BFA models was trained on any of the three corpora (Table~\ref{tab:corpora}). WildPhonemes, whose reference BFA helped to seed, serves only as a check and is described in Section~\ref{sec:res-wild}. No reference was drawn from scratch: each started from an automatic alignment that annotators then corrected, which matters for how the systems compare (Section~\ref{sec:aligner_bias}).

\subsection{Metric}\label{sec:metric}

For each segment, predicted onsets of speech phones and reference onsets are matched one-to-one, in time order, by an edit-distance alignment in which two onsets can be paired only if they lie within 100\,ms of each other. The limit separates imprecise boundaries from gross errors. An onset more than 100\,ms from its reference, about the length of a phone, is not a badly timed version of that boundary but a different mistake, and it counts as a miss and a false alarm instead of entering the error statistics. A match within the tolerance $\tau$ is a hit, from which precision, recall and F1 follow. Labels are not compared: a correctly timed boundary between wrongly named phones still counts, which tests timing fairly for systems with different phone sets. Because matching is one-to-one, extra predicted onsets cannot stack on one boundary to inflate recall, and each unmatched one lowers precision.

We report F1 at 10, 25 and 50\,ms (\Fone{10} and so on), recall and precision at 25\,ms, the median absolute error of matched onsets, and the ratio of predicted to reference onsets. The tight tolerances match human agreement: in the studies surveyed by \citet{young2023forced}, two labellers placed a little over half to four fifths of boundaries within 10\,ms of each other and most within 20\,ms. They also match the cost of hand correction, since moving a boundary that is slightly off takes about as long as moving one that is far off \citep{young2023forced}, so what saves annotation time is the share of boundaries that need no correction. They matter, too, because a per-boundary score at a loose one flatters an aligner: if 95\% of the boundaries are within 50\,ms and errors are independent, an utterance of 20 phones still has at least one boundary beyond 50\,ms about two times in three ($1 - 0.95^{20} = 0.64$). As a lenient view we also report, for each reference boundary, the distance to the nearest predicted onset. Any single tolerance is arbitrary \citep{deng2025speech}; the three tolerances, the median error and the nearest-onset curve over all tolerances (Fig.~\ref{fig:nearest}) are reported together for that reason.

\subsection{Systems}\label{sec:systems}

\paragraph{BFA.} The six models that cover English: CUPE-3-en, CUPE-3-en-base and CUPE-3-mls7 (ph66), and mbfa-latin, mbfa-latin-benchmark and mbfa-latin-noEnglish (std). All run at the decoder defaults, with frame-grid onsets and batches of 16 segments.

\paragraph{CUPE-2i with CTC.} Our earlier CTC baseline: the CTC-trained CUPE-2i English phoneme model \citep{rehman2025cupe}\footnote{\url{https://huggingface.co/Tabahi/CUPE-2i}} (960\,h of LibriSpeech, 13.125\,ms frames) with basic CTC decoding of its phone posteriors, a Viterbi search over the target phones with a blank class, as released in BFA v1.1.5 \citep{rehman2026bfa1x}, with that release's optional onset snapping off. Like the CUPE-3 models it has no long-range context: it sees no audio beyond a 120\,ms window. It aligns the same espeak-ng ph66 transcripts as the CUPE-3 models. On Buckeye it returns no phones for 234 segments and is scored on 99.5\% of the boundaries.

\paragraph{wav2vec2 XLSR-53 with CTC.} A CTC baseline that is not our own: the wav2vec~2.0 XLSR-53 model \citep{conneau2021xlsr} fine-tuned with CTC on eSpeak phoneme transcripts of multilingual Common Voice \citep{xu2022zeroshot}\footnote{\url{https://huggingface.co/facebook/wav2vec2-xlsr-53-espeak-cv-ft}} (20\,ms frames), aligned by the CTC forced alignment of torchaudio \citep{hwang2023torchaudio}, a Viterbi search over the target tokens with a blank class. Neither TIMIT nor Buckeye is in its training data. Its phone set is eSpeak's, so it aligns the same espeak-ng ph66 transcripts as the CUPE-3 models, each phone given as its eSpeak token. The few eSpeak symbols that ph66 splits into two phones, such as r-coloured vowels, are given as two tokens. A phone's onset is the first frame of its span on the Viterbi path, with no shift or other post-processing. We do not run word-level CTC aligners such as MMS and WhisperX: they time words rather than phones.

\paragraph{Charsiu W2V2-FS.} The published neural aligner closest to ours in supervision: Charsiu's W2V2-FS \citep{zhu2022charsiu}\footnote{\url{https://huggingface.co/charsiu/en_w2v2_fs_10ms}}, which learns the alignment without boundary labels. A wav2vec~2.0 base encoder with 10\,ms frames and a phone encoder are joined by one cross-attention layer, trained with wav2vec~2.0's contrastive loss and a forward-sum loss that keeps the attention monotonic. It was trained in four rounds, on Common Voice, then LibriSpeech, and in the last round on the audio and transcripts of the TIMIT training set. The released checkpoint does not say which round it is, so we assume that it saw the TIMIT training set, which holds 4,620 of the 6,300 TIMIT clips we score. Its TIMIT row in Table~\ref{tab:main} is therefore marked, and Section~\ref{sec:main} reports the 1,680 test clips on their own. It never saw Buckeye. It uses its own front end, g2p\_en \citep{park2019g2pe} with the 39 ARPAbet phones of the CMU dictionary and silence at both ends of each segment, and is scored through the same ARPAbet table as MFA. A phone's onset is the first frame of its span on the best monotonic path through the attention matrix, which keeps every phone in transcript order. Charsiu's released code instead gives each frame its most attended phone, which can repeat or skip phones, and scores no better. We run the network in evaluation mode. 

\paragraph{MAPS.} The Mason-Alberta Phonetic Segmenter \citep{kelley2024maps}\footnote{\url{https://github.com/MasonPhonLab/MAPS}}: a BiLSTM over MFCCs with 10\,ms frames gives TIMIT phone posteriors, a forced alignment over them places the phones, and a linear interpolation of the posteriors places each boundary inside its frame. We run its released English model through its own code, with MFA's \texttt{english\_us\_arpa} dictionary and MFA's G2P for the words that dictionary lacks; it is scored through the same ARPAbet table as MFA. Unlike every other system here, it was trained on hand-labelled boundaries, and they were the boundaries of TIMIT's training set and of all of Buckeye except four speakers (s27, s38, s39 and s40), so it has learned the labelling conventions of both corpora. Its rows in Table~\ref{tab:main} are therefore scored only on the clips it did not see, the 1,680 clips of TIMIT's test set and the 25 recordings of those four speakers. They are marked, are not bolded, and the table note gives the other systems' scores on the same clips. Even on these clips MAPS is an in-domain system, with the recording conditions and labelling conventions of its training data. MAPS and Charsiu W2V2-FS are marked for different reasons: MAPS learned the boundaries of TIMIT and Buckeye, Charsiu only the audio and transcripts of TIMIT's training set.

\paragraph{MFA.} MFA 3.4.3 \citep{mcauliffe2017mfa} with two pretrained English set-ups: the \texttt{english\_us\_arpa} acoustic model and dictionary (ARPAbet), and the \texttt{english\_mfa} acoustic model with the \texttt{english\_us\_mfa} dictionary (IPA, with more pronunciation variants). We use the default settings, with speaker adaptation on. Each speaker's segments are in their own folder, so that speaker adaptation is per speaker; a flat folder would make the whole corpus one speaker and costs MFA up to about one point of \Fone{25} (Supplementary Section~S2). On Buckeye both models return no phones for a few segments; they are scored on the remaining 99.8\% of the boundaries.

\paragraph{Not compared.} Charsiu's frame classifier, W2V2-FC, is trained on MFA alignments of LibriSpeech, so its boundaries are MFA's; we do not run it. MAUS \citep{kisler2017maus} is not compared either: the corpora its acoustic models were trained on are not documented clip by clip, so, unlike MAPS, it cannot be scored on clips it is known not to have seen.

\section{Results}\label{sec:results}

The results are organised around one question: how well does an aligner with no dictionary and no boundary labels place boundaries in English, including when English was left out of its training? Section~\ref{sec:res-wild} adds a check of the models for the other script groups, against a reference that is human-verified but partial to BFA. MFA and the neural aligners are run alongside BFA as reference points for what a given score means, not as competitors in a ranking.

\subsection{Boundary accuracy on English}\label{sec:main}

TIMIT and Buckeye are the only corpora here whose phone boundaries were placed by hand, so they are where boundary precision can be measured most directly.

\begin{table}[t]
\caption{Full-corpus results. F1, recall (R) and precision (P) in \%, at the tolerance given; median absolute error of matched onsets; ratio of predicted to reference onsets. Systems as in Section~\ref{sec:systems}, BFA at its defaults. \textsuperscript{\textdagger}\textsuperscript{\textdaggerdbl}Not like-for-like comparisons: see the notes below the table. Best per column in bold, not counting the \textsuperscript{\textdaggerdbl} rows, which are scored on held-out clips only.}\label{tab:main}
\small
\begin{threeparttable}
\begin{tabular*}{\tblwidth}{@{\extracolsep{\fill}}llrrrrrrr@{}}
\toprule
System & Set & \Fone{10} & \Fone{25} & \Fone{50} & R@25 & P@25 & Med.\ err. & Pred/ref \\
\midrule
\multicolumn{9}{@{}l}{\emph{TIMIT (read speech, hand-labelled)}} \\
BFA CUPE-3-en & ph66 & \textbf{59.9} & 78.2 & 87.9 & 75.5 & 81.0 & \textbf{7\,ms} & 0.93 \\
BFA CUPE-3-en-base & ph66 & 59.5 & 78.6 & 88.2 & 75.9 & 81.5 & \textbf{7\,ms} & 0.93 \\
BFA CUPE-3-mls7 & ph66 & 55.2 & 73.1 & 84.5 & 70.6 & 75.8 & 8\,ms & 0.93 \\
BFA mbfa-latin & std & 54.5 & 73.6 & 85.3 & 73.2 & 74.0 & 8\,ms & 0.99 \\
BFA mbfa-latin-benchmark & std & 56.2 & 75.2 & 86.1 & 74.9 & 75.6 & 8\,ms & 0.99 \\
BFA mbfa-latin-noEnglish & std & 54.4 & 73.2 & 84.6 & 72.9 & 73.6 & 8\,ms & 0.99 \\
CUPE-2i with CTC & ph66 & 22.8 & 50.1 & 77.3 & 48.4 & 52.0 & 23\,ms & 0.93 \\
wav2vec2 XLSR-53 with CTC & ph66 & 27.6 & 57.6 & 83.7 & 55.6 & 59.7 & 19\,ms & 0.93 \\
Charsiu W2V2-FS\tnote{\textdagger} & ARPAbet & 40.2 & 76.1 & \textbf{91.1} & 73.1 & 79.3 & 13\,ms & 0.92 \\
MAPS\tnote{\textdaggerdbl} & ARPAbet & 58.9 & 80.3 & 88.4 & 77.0 & 83.8 & 7\,ms & 0.92 \\
MFA english\_us\_arpa & ARPAbet & 55.2 & \textbf{80.9} & 90.4 & \textbf{77.0} & \textbf{85.1} & 8\,ms & 0.90 \\
MFA english\_mfa & IPA & 56.5 & 78.6 & 89.1 & 73.8 & 84.0 & 8\,ms & 0.88 \\
\midrule
\multicolumn{9}{@{}l}{\emph{Buckeye (conversational, HMM-seeded labels)}} \\
BFA CUPE-3-en & ph66 & 53.5 & 76.5 & 88.2 & 81.1 & 72.4 & 9\,ms & 1.12 \\
BFA CUPE-3-en-base & ph66 & 53.3 & 76.4 & 88.1 & 81.0 & 72.4 & 9\,ms & 1.12 \\
BFA CUPE-3-mls7 & ph66 & 48.5 & 70.3 & 83.5 & 74.5 & 66.6 & 10\,ms & 1.12 \\
BFA mbfa-latin & std & 50.0 & 71.2 & 82.8 & 78.5 & 65.2 & 9\,ms & 1.20 \\
BFA mbfa-latin-benchmark & std & 50.8 & 72.0 & 83.3 & 79.4 & 65.9 & 9\,ms & 1.20 \\
BFA mbfa-latin-noEnglish & std & 49.8 & 70.9 & 82.5 & 78.1 & 64.8 & 9\,ms & 1.20 \\
CUPE-2i with CTC & ph66 & 25.6 & 52.3 & 77.1 & 54.7 & 50.2 & 20\,ms & 1.09 \\
wav2vec2 XLSR-53 with CTC & ph66 & 25.4 & 54.9 & 81.8 & 58.2 & 52.0 & 20\,ms & 1.12 \\
Charsiu W2V2-FS & ARPAbet & 43.8 & 75.6 & 88.5 & 79.1 & 72.4 & 11\,ms & 1.09 \\
MAPS\tnote{\textdaggerdbl} & ARPAbet & 47.7 & 74.5 & 85.3 & 77.4 & 71.9 & 10\,ms & 1.08 \\
MFA english\_us\_arpa & ARPAbet & 60.6 & 84.3 & 92.1 & \textbf{87.3} & 81.5 & 8\,ms & 1.07 \\
MFA english\_mfa & IPA & \textbf{63.6} & \textbf{85.3} & \textbf{93.2} & 85.9 & \textbf{84.7} & \textbf{7\,ms} & 1.01 \\
\bottomrule
\end{tabular*}
\begin{tablenotes}\footnotesize
\item[\textdagger] Charsiu W2V2-FS was fine-tuned on the audio and transcripts (not the boundaries) of the TIMIT training set, 4,620 of the 6,300 clips scored here; no BFA model saw TIMIT or Buckeye. On the 1,680 TIMIT test clips alone it scores 40.0 / 76.3 / 91.2 (\Fone{10} / \Fone{25} / \Fone{50}), against 59.8 / 78.0 / 87.9 for BFA CUPE-3-en on the same clips.
\item[\textdaggerdbl] MAPS was trained on the hand-labelled boundaries of TIMIT's training set and of Buckeye except four speakers. Its rows are scored only on the clips it did not see: TIMIT's 1,680 test clips and the four held-out Buckeye speakers (25 recordings). On the same clips BFA CUPE-3-en scores 59.8 / 78.0 / 87.9 on TIMIT and 52.3 / 75.9 / 88.1 on Buckeye, and MFA english\_us\_arpa 55.1 / 81.2 / 90.7 and 60.2 / 83.1 / 91.6 (\Fone{10} / \Fone{25} / \Fone{50}).
\end{tablenotes}
\end{threeparttable}
\end{table}

\begin{figure}[pos=!htb]
  \centering
  \includegraphics[width=\textwidth]{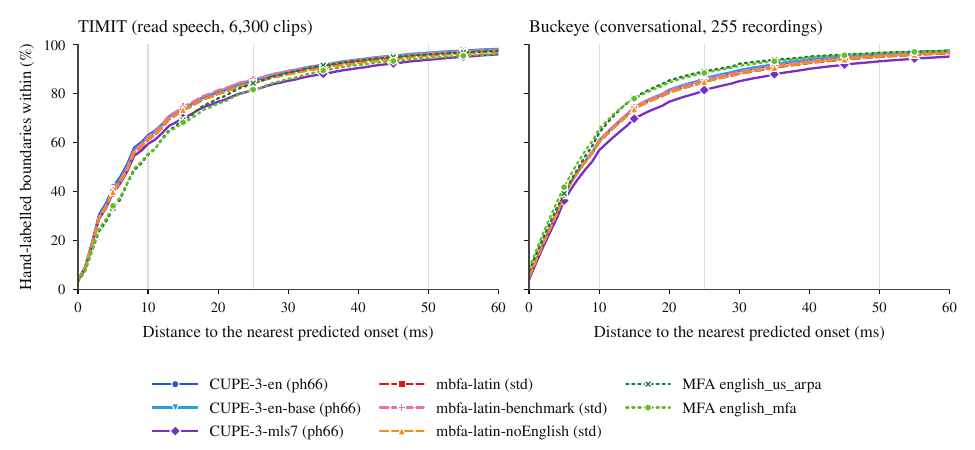}
  \caption{Share of reference boundaries with a predicted onset within $t$\,ms, per system (BFA at its defaults: ph66 models solid, std models dashed; MFA, with one folder per speaker, dotted). Left: TIMIT, all 6,300 clips. Right: Buckeye, all 255 recordings.}
  \label{fig:nearest}
\end{figure}

\paragraph{Read speech.} On TIMIT (Table~\ref{tab:main}), the English-trained BFA models place most onsets within 25\,ms of the hand labels. The nearest-onset curve (Fig.~\ref{fig:nearest}) shows, at every tolerance, how many reference boundaries each system comes close to, whether or not its other onsets are right. BFA reads its boundaries on a 5\,ms grid, and rounding them to a 10\,ms grid costs about three points of \Fone{10} (Section~\ref{sec:res-trail}). When BFA is wrong it tends to be wrong by more, and it places onsets that the reference does not have; the median error ignores such errors, while a one-to-one F1 at 25\,ms counts them. TIMIT also caps recall for every system alike: it labels stops as closure plus release, and about one reference boundary in ten is a release that no system placing one onset per stop from text can match.

\paragraph{Conversational speech.} On Buckeye, BFA loses more in precision than in recall: it places more onsets than the reference has (Pred/ref in Table~\ref{tab:main}). Two likely causes are canonical pronunciations of speech that the speaker reduced, and a reference that was seeded by an HMM aligner (Section~\ref{sec:aligner_bias}).

\paragraph{CTC decoding.} The two CTC rows align the same ph66 transcripts as the CUPE-3 models and place as many onsets, so the phone sequence is held fixed. Both put most boundaries roughly in the right place, as their \Fone{50} shows, but few of them within 10\,ms. This is consistent with CTC's peaky posteriors, which leave the boundary inside a blank run to the decoder (Section~\ref{sec:intro}). The frame rates explain only part of it: rounding to a 13.125 or 20\,ms grid moves an onset by at most 6.6 or 10\,ms, less than either system's median error, and rounding BFA's onsets to the same grids costs it under one point of \Fone{25} (Section~\ref{sec:res-trail}). Encoder, training data, frame rate and decoder all differ from the CUPE-3 models, so these rows show what CTC decoding of phone posteriors gives in practice, not a limit of the CTC objective. Section~\ref{sec:res-trail} traces the steps from CUPE-2i to CUPE-3, among them a CTC run and a hat-target run matched in data and training length.

\paragraph{Other neural aligners.} Charsiu W2V2-FS, the other aligner here that learns alignment without boundary labels, also places most boundaries in the right place but fewer of them within 10\,ms. Its 10\,ms frames do not explain this, since MFA places its boundaries on the same grid. Its TIMIT row is marked because it was fine-tuned on the audio and transcripts of TIMIT's training set, but it scores the same on the test clips as on the training clips, so this exposure did not raise its score (note to Table~\ref{tab:main}). Charsiu's wav2vec~2.0 encoder sees the whole segment, and so does wav2vec2 XLSR-53 under CTC decoding, whereas the CUPE-3 models see 39\,ms of audio per frame; long context is therefore not what makes boundaries sharp, although these rows differ in more than one respect at a time. MAPS, trained on the hand-labelled boundaries of TIMIT and Buckeye, shows what supervision with a corpus's own labels gives on that corpus's held-out speakers (note to Table~\ref{tab:main}).

\subsection{Aligning English without training for it}\label{sec:res-unseen}

Two BFA models never heard English: CUPE-3-mls7, trained from random weights on seven MLS languages, and mbfa-latin-noEnglish, trained from random weights on the Latin-script FLEURS data with English removed (35 languages). Both align TIMIT and Buckeye from G2P pronunciations, with no English dictionary and no English audio.

\paragraph{English is not needed.} mbfa-latin-noEnglish comes within half a point of mbfa-latin, which heard 7\,h of English among its 36 languages, on both corpora at 10 and 25\,ms (Table~\ref{tab:main}). The two Latin-script runs are not a matched pair, so the result shows that English training data are not needed for this accuracy, not that they have no effect. Removing English also barely changes what the model is trained on: English is 2.7\% of the Latin-script training phones, its phones are with few exceptions also phones of the other 35 languages, and a network that sees 39\,ms of audio has no access to the words, phonotactics or prosody of English.

\paragraph{Language-specific data still help.} CUPE-3-en is CUPE-3-mls7 continued on LibriSpeech, and is more accurate on both corpora; mbfa-latin, for which English is under 3\% of the data, is less accurate than the English-dominant CUPE-3-en. That gap lies either in the acoustic model, trained on 273\,h spread over 36 languages rather than about 860\,h of English, or in the pronunciations, since the two families use different G2P front ends. We did not separate the two. Training data in the target language improve accuracy, but they are not required to align a language whose phones the model has heard in others.

\subsection{A check in fourteen languages: WildPhonemes}\label{sec:res-wild}

TIMIT and Buckeye cover only English. To check the models of the other script groups, we used WildPhonemes \citep{ar2026_wildphonemes}, a word- and phone-aligned set that we drew from Common Voice 22.0 \citep{ardila2020commonvoice} in 14 languages, at least one per script group (Table~\ref{tab:wildphonemes}). Its reference was made as TIMIT's and Buckeye's were, by correcting an automatic alignment by hand, but that alignment was in part BFA's own, and the set is new rather than an established benchmark. Its scores are therefore partial to BFA, and we use them only to verify that the script-group models work in their languages, as a sanity check, not to compare BFA with other aligners. Clips come from the test split (also dev and train for ko, vi and yue, whose test pools are small) and had to pass every filter: no down-votes and at least two up-votes; 5--30\,s long; only the language's own script and nothing that needs expanding, such as digits or symbols; a prompt within 5\% CER of a Whisper large-v3 \citep{radford2023whisper} transcription (8--10\% for languages where Whisper itself is weaker); a G2P output without unknown tokens; DNSMOS P.835 \citep{reddy2022dnsmos} BAK $\ge 3.3$ and SIG $\ge 3.0$, no clipping and no internal pause over 1.5\,s. Clips were drawn round-robin over speakers, alternating gender where Common Voice records it, up to 100 clips or 10\,min per language. Listeners discarded 12 of the 1,215 clips drawn, which leaves the set of Table~\ref{tab:corpora}, with 14,939 words.

\paragraph{Reference.} Each word edge started at the midpoint of the BFA and MFA \citep{mcauliffe2017mfa} alignments (from BFA alone for about 1,500 edges that MFA did not pair). The authors visually checked every edge on the spectrogram and waveform with no aligner information shown, and placed by hand every edge on which the two aligners differed by more than 100\,ms. The phones inside each word were then force-aligned between its two fixed edges with BFA's acoustic model, in BFA's std phone set. Finally, two annotators listened to every clip, checked the word edges and the inner phone boundaries, and moved those that were misplaced, with closer attention to word junctions with weak cues, possible transcript mismatches and regions noted in the visual pass. English and Mandarin had a third annotation pass by the authors. For all other languages, the mean of the two annotators' placements was taken as the final boundary, and disagreements of more than 50\,ms were resolved by the authors. In rare cases errors were left in, such as word insertions or deletions where forced alignment still had to place every word; these are rare except in French. Like TIMIT and Buckeye, the set is human-verified rather than labelled from scratch, since automatic seeding keeps a precision that boundaries placed by ear alone do not reach; that precision can be partly verified from obvious changes in the spectrum, such as the boundaries between stops and vowels. Aligners are also precise at marking boundaries that are clearly visible in the spectrum. Most mistakes happen when a whole word (or several) is misplaced, which is why we realigned the phone boundaries inside each word only after verifying its edges. BFA therefore seeded every inner phone boundary and, with MFA, every word edge, and wherever annotators accepted the seed the reference keeps BFA's placement (Section~\ref{sec:aligner_bias}). The verification changed a large share of the seed, at both levels: in the final reference, 46.4\% of the phone onsets and 69.4\% of the word edges lie more than 5\,ms from the nearest corresponding boundary of BFA's alignment, and every clip contains such onsets and edges.

\paragraph{Metrics and systems.} We score three kinds of reference boundary. \emph{Word edges} are paired through the text, not through time: both word lists are reduced to letters and aligned character by character, and each reference edge is paired with the predicted edge at the same character, so the score depends on neither aligner's phone set; we report the share within 25\,ms. \emph{All phone onsets} are scored by \Fone{25} as in Section~\ref{sec:metric}. \emph{Moved onsets} are the reference onsets more than 5\,ms from the nearest onset of BFA's alignment, the 32,655 onsets on which the reference departs from BFA; they are matched as all onsets are, but only recall is defined on this subset, and we report the share within 50\,ms. BFA runs the released \texttt{mbfa-<group>} model of each language at its defaults, and CUPE-3-en runs on en. MFA~3 is run alongside as a reference point for the scale of the scores. It uses the pretrained \texttt{*\_mfa} acoustic model and dictionary where one exists; for el and ta it uses MFA's released Common Voice models and dictionaries (\texttt{greek\_cv}, \texttt{tamil\_cv}), and for ar, ur and yue, which have neither, we trained a model on 5.5--7.2\,h of that language's Common Voice data, including the clips of this set, with a dictionary from BFA's G2P. Every MFA model used here except Korean's includes Common Voice in its training data \citep{mfamodels}: the pretrained ones as part of a much larger and more varied collection, the three we trained as their only source. Every BFA model was trained on FLEURS (Section~\ref{sec:data}).

\begin{table}[t]
\caption{WildPhonemes check: share of boundaries within the tolerance (\%), BFA with MFA as a reference point (Section~\ref{sec:res-wild}). BFA seeded the reference (every inner phone boundary and, with MFA, every word edge), so these scores are partial to BFA: they verify the models but do not rank the aligners, and no column is bolded. Means unweighted over languages. \textsuperscript{\textdaggerdbl}Only 223 of 3,708 ja word edges can be paired with MFA's word segmentation (zh-CN: 2,140 of 3,110). CUPE-3-en on en: 65.5 / 73.5 / 79.7.}\label{tab:wildphonemes}
\small
\setlength{\tabcolsep}{2.7pt}
\setlength{\aboverulesep}{0pt}\setlength{\belowrulesep}{0pt}%
\renewcommand{\arraystretch}{1.15}
\begin{tabular}{@{}llrrr*{2}{>{\centering\arraybackslash}p{39.5pt}}*{2}{>{\centering\arraybackslash}p{39.5pt}}*{2}{>{\centering\arraybackslash}p{39.5pt}}@{}}
\toprule
\rule[-1.1ex]{0pt}{0pt} & & & & \multicolumn{1}{r|}{Onsets} & \multicolumn{2}{c|}{\begin{tabular}[b]{@{}c@{}}Word edges\\$\le$25\,ms\end{tabular}} & \multicolumn{2}{c|}{\begin{tabular}[b]{@{}c@{}}All onsets\\\Fone{25}\end{tabular}} & \multicolumn{2}{c}{\begin{tabular}[b]{@{}c@{}}Moved onsets\\$\le$50\,ms\end{tabular}} \\
\cmidrule(lr){6-7}\cmidrule(lr){8-9}\cmidrule(l){10-11}
Lang\rule{0pt}{2.6ex}\rule[-1.1ex]{0pt}{0pt} & Group & Clips & Min & \multicolumn{1}{r|}{all / moved} & BFA & \multicolumn{1}{c|}{MFA} & BFA & \multicolumn{1}{c|}{MFA} & BFA & MFA \\
\midrule
en    & latin        & 95 & 9.9  & 5,086 / 2,267 & 73.3 & 74.8 & 82.3 & 68.5 & 87.2 & 78.9 \\
es    & latin        & 97 & 10.0 & 5,946 / 2,525 & 82.0 & 81.2 & 91.7 & 75.0 & 93.2 & 92.2 \\
fr    & latin        & 96 & 9.9  & 4,573 / 2,172 & 72.1 & 76.5 & 85.0 & 68.1 & 84.0 & 84.7 \\
ru    & cyrillic     & 95 & 9.9  & 7,169 / 3,121 & 83.5 & 84.6 & 90.4 & 72.4 & 94.1 & 90.6 \\
el    & other\_alph. & 94 & 9.2  & 4,446 / 2,166 & 75.4 & 72.8 & 87.1 & 70.8 & 91.0 & 88.6 \\
ko    & other\_alph. & 61 & 6.3  & 3,987 / 1,840 & 83.8 & 85.4 & 88.6 & 69.3 & 90.0 & 83.9 \\
vi    & vietnamese   & 24 & 2.2  & 756 / 388     & 69.5 & 69.3 & 84.8 & 66.3 & 86.9 & 85.8 \\
th    & thai\_khmer  & 98 & 10.0 & 4,713 / 2,056 & 83.7 & 86.5 & 90.0 & 77.6 & 90.9 & 89.3 \\
ta    & brahmic      & 83 & 9.1  & 5,535 / 2,485 & 55.5 & 45.7 & 85.8 & 62.6 & 85.2 & 70.6 \\
ar    & abjad        & 98 & 9.8  & 6,100 / 2,700 & 73.0 & 72.7 & 87.5 & 61.2 & 87.8 & 81.0 \\
ur    & abjad        & 97 & 9.8  & 5,137 / 2,263 & 81.5 & 82.5 & 88.7 & 72.0 & 90.8 & 89.8 \\
ja    & japanese     & 82 & 9.7  & 6,624 / 3,097 & 81.2 & 70.4\textsuperscript{\textdaggerdbl} & 86.9 & 74.3 & 91.0 & 88.3 \\
yue   & cjk          & 94 & 9.9  & 4,998 / 2,719 & 73.9 & 86.9 & 80.9 & 74.0 & 88.3 & 93.2 \\
zh-CN & cjk          & 89 & 9.9  & 5,286 / 2,856 & 68.7 & 82.7 & 78.1 & 70.2 & 85.3 & 75.5 \\
\midrule
\multicolumn{2}{@{}l}{Total / mean} & 1,203 & 125.6 & 70,356 / 32,655 & 75.5 & 76.6 & 86.3 & 70.2 & 89.0 & 85.2 \\
\bottomrule
\end{tabular}
\end{table}

\paragraph{What the check shows.} In every language, most of the word edges placed by the script-group models lie within 25\,ms of the verified reference, and MFA's scores on the same reference are of the same order (Table~\ref{tab:wildphonemes}). No language shows the kind of failure that a wrong token list, a broken G2P or an unsuitable adaptation would cause. This holds although every non-Latin model was adapted from the Latin-script model on far less audio, under 10\,h for four of them (Table~\ref{tab:models}), was trained on FLEURS rather than Common Voice, and was decoded with defaults chosen on English. The check also shows where the models are weakest: on the word edges of Tamil, Mandarin and Vietnamese. Few of MFA's Japanese word edges can be paired, so its Japanese column says little.

\paragraph{Phone onsets.} The scores over all phone onsets say little about BFA: the inner phones of the reference were placed with BFA's acoustic model, in its phone set, so the reference favours BFA by construction, and MFA cannot hit the second token of BFA's two-token diphthongs and affricates. The moved onsets were placed by the verification more than 5\,ms from BFA's alignment, so they guard against reading BFA's own seed back. Their scores show mainly how far the verification moved them, not how precise BFA is.

Together with the English results, the check supports using the script-group models in the languages they were trained on. Their accuracy is human-verified but on a reference partial to BFA; an impartial measurement needs a reference that no aligner under test helped to seed (Section~\ref{sec:limitations}).

\subsection{Robustness to transcript errors}\label{sec:res-errors}

A forced aligner is only as good as its transcript. Because forced alignment places every phone it is given, a phone that is not in the audio has to take time from its neighbours, and a phone missing from the text leaves its audio to them. We measure how much this costs each system, and how far the damage spreads.

\paragraph{Protocol.} We corrupt the transcripts of TIMIT and Buckeye and leave the audio untouched, so every condition is scored against the same hand labels; this is an accuracy test. At the \emph{phone level}, every $n$-th phone of each clip, from a random offset seeded per clip and at the same sites for every system, is \emph{deleted}, preceded by an \emph{inserted} unspoken phone, substituted by a phone of the \emph{same} broad group (e.g.\ /t/ $\to$ /d/), or substituted by one of the \emph{other} major class (consonant $\leftrightarrow$ vowel). Replacements are drawn from the corpus' own phones in each system's phone set. At the \emph{word level}, closer to ASR errors, every $n$-th word is deleted, preceded by an inserted word, or replaced by another word drawn from the corpus vocabulary by the same seeded rule for every system; each system writes the words in its own phone set.

BFA receives the corrupted phone sequence in place of its phonemiser's output, and Charsiu W2V2-FS in place of g2p\_en's phones, between its two silence tokens. MFA receives it as one private dictionary entry per word, built from its own most likely pronunciation, and so does MAPS, from the english\_us\_arpa pronunciation it uses in Table~\ref{tab:main}. Its clean condition uses the same private dictionary, so the only handicap this imposes, the loss of alternative pronunciations, is common to every condition. Predicted onsets of every phone, inserted ones included, are scored as in Section~\ref{sec:metric}.

We use every clip whose words both MFA dictionaries can spell, a contraction missing from a dictionary being spelled as its base and clitic: 5,568 of TIMIT's 6,300 clips and 32,545 of Buckeye's 33,676 segments of at most 30\,s. On Buckeye, 23 of these segments run past the end of their audio and are skipped by every system, and MFA returns no alignment for 114 or more. MAPS learned the boundaries of TIMIT's training set and of all but four Buckeye speakers (Section~\ref{sec:systems}), so it is scored only on the clips it did not see: the 1,498 TIMIT test clips and the 3,144 segments of the four held-out speakers in this set. Every other system's full run is also scored on these clips, and each one's mean drop in \Fone{25} at every rate is within 0.4 points of its drop on the full corpus, so MAPS's drops can be read against the others'. Phones are corrupted at rates of 5, 10 and 20\%, words at 10 and 20\%. BFA runs at its defaults throughout (CUPE-3-en, CUPE-3-mls7, mbfa-latin and mbfa-latin-noEnglish), and also at other onset weights $\beta$ to isolate the onset term.

\begin{figure}[pos=!htb]
  \centering
  \includegraphics[width=\textwidth]{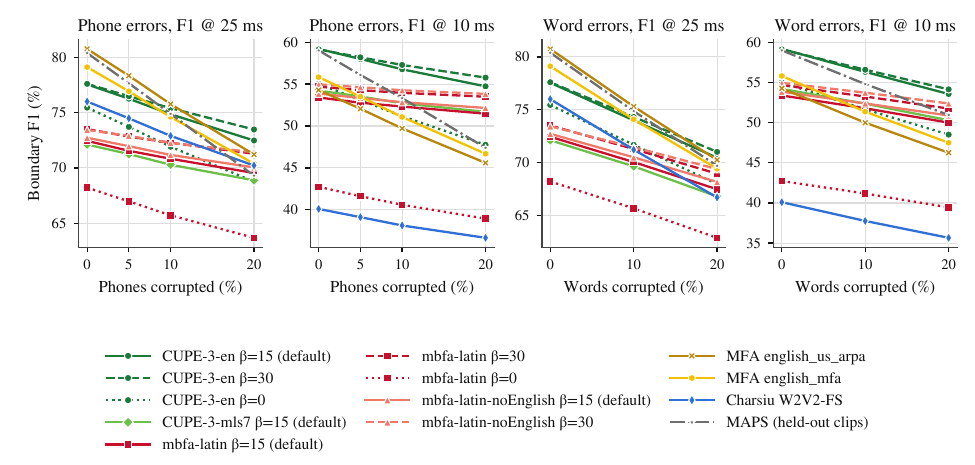}
  \caption{\Fone{25} and \Fone{10} on TIMIT against the share of phones (left two panels) or words (right two panels) corrupted, scored against the unchanged hand labels. Each point is the mean over the error kinds at that rate; 0\% is the clean transcript. Colour is the model family (CUPE green, mbfa red, MFA yellow, Charsiu blue, MAPS grey), with one shade and marker per model. BFA is shown at its default onset weight $\beta = 15$ (solid), at $\beta = 30$ (dashed) and without the onset term ($\beta = 0$, dotted). MAPS (dash-dot) is scored on the 1,498 TIMIT test clips, every other system on all 5,568 clips. Buckeye and WildPhonemes English are summarised in Table~\ref{tab:errslope}.}
  \label{fig:errors}
\end{figure}

As a check outside the two corpora, we repeat the test on the 95 English clips of WildPhonemes (Section~\ref{sec:res-wild}), which no system was trained on, MAPS included; the few words missing from an MFA dictionary are spelled by MFA's G2P. Its reference was made by correcting an alignment that BFA seeded, so these scores favour BFA, and we report them only beside the hand-labelled results. Because every curve is close to a straight line (Fig.~\ref{fig:errors}), we summarise each system on each corpus by its \emph{slope}: the least-squares fit of \Fone{25} against the error rate, in points per 10\% of phones or words corrupted (Table~\ref{tab:errslope}).

\begin{table}[t]
\caption{Transcript errors at 10\% of phones. \Fone{25} with \Fone{10} in brackets; ``worst drop'' is the largest fall in \Fone{25} from the clean transcript. Best per corpus and column in bold, not counting MAPS. \textsuperscript{\textdagger}Saw the text of TIMIT's training set (Table~\ref{tab:main}). \textsuperscript{\textdaggerdbl}Trained on TIMIT's training set and on Buckeye but four speakers: scored on its held-out clips instead (1,498 TIMIT test clips; 3,144 Buckeye segments), where CUPE-3-en at $\beta = 15$ scores 77.6 (59.1) clean on TIMIT and 76.0 (52.6) on Buckeye and loses at most 3.5 and 4.5 points. Every rate and the word-level errors are in Supplementary Section~S3.}\label{tab:errors}
\small
\begin{tabular*}{\tblwidth}{@{\extracolsep{\fill}}lllllll@{}}
\toprule
System & Clean & Delete & Insert & Similar sub. & Other-class sub. & Worst drop \\
\midrule
\multicolumn{7}{@{}l}{\emph{TIMIT}} \\
CUPE-3-en, $\beta = 15$ (default) & 77.6 (\textbf{59.2}) & 74.2 (57.1) & 74.5 (55.7) & 76.0 (57.7) & 74.7 (56.5) & $-3.3$ \\
CUPE-3-en, $\beta = 30$ & 77.6 (59.1) & \textbf{74.8} (\textbf{57.7}) & 75.0 (\textbf{56.1}) & 76.5 (\textbf{58.2}) & \textbf{75.4} (\textbf{57.2}) & $-2.8$ \\
CUPE-3-en, $\beta = 0$ & 75.5 (55.0) & 70.9 (51.1) & 72.2 (50.9) & 73.3 (52.5) & 71.3 (50.2) & $-4.5$ \\
CUPE-3-mls7, $\beta = 15$ & 72.1 (54.2) & 69.5 (52.9) & 70.2 (51.9) & 71.4 (53.6) & 70.2 (52.5) & $-2.6$ \\
mbfa-latin, $\beta = 15$ (default) & 72.5 (53.4) & 70.1 (52.5) & 71.0 (51.9) & 71.8 (52.9) & 70.5 (52.0) & $-2.3$ \\
mbfa-latin, $\beta = 30$ & 73.5 (54.7) & 71.7 (54.4) & 72.3 (53.3) & 73.0 (54.4) & 72.1 (53.8) & $\mathbf{-1.8}$ \\
mbfa-latin-noEnglish, $\beta = 15$ & 72.7 (53.9) & 70.6 (53.2) & 71.2 (52.2) & 72.1 (53.4) & 70.8 (52.5) & $-2.1$ \\
Charsiu W2V2-FS\textsuperscript{\textdagger} & 76.0 (40.1) & 71.3 (37.5) & 71.9 (37.4) & 75.6 (39.8) & 72.8 (37.9) & $-4.7$ \\
MAPS\textsuperscript{\textdaggerdbl} & 80.4 (59.0) & 74.0 (52.6) & 76.2 (55.2) & 76.3 (54.0) & 73.1 (51.5) & $-7.3$ \\
MFA english\_us\_arpa & \textbf{80.8} (54.3) & 74.3 (49.0) & \textbf{75.3} (49.1) & \textbf{79.9} (53.5) & 73.6 (47.3) & $-7.2$ \\
MFA english\_mfa & 79.1 (55.8) & 73.1 (50.2) & 74.6 (50.8) & 78.1 (54.5) & 72.7 (48.8) & $-6.4$ \\
\midrule
\multicolumn{7}{@{}l}{\emph{Buckeye}} \\
CUPE-3-en, $\beta = 15$ (default) & 76.5 (53.5) & 71.9 (49.7) & 74.7 (52.5) & 75.0 (52.1) & 73.2 (50.6) & $-4.6$ \\
CUPE-3-en, $\beta = 30$ & 75.1 (52.1) & 71.1 (49.0) & 73.7 (51.5) & 73.9 (51.1) & 72.3 (49.8) & $-4.0$ \\
CUPE-3-en, $\beta = 0$ & 77.7 (52.4) & 71.8 (46.7) & 75.3 (50.8) & 75.6 (50.0) & 73.3 (47.7) & $-5.9$ \\
CUPE-3-mls7, $\beta = 15$ & 70.3 (48.5) & 66.7 (45.5) & 69.4 (48.1) & 69.3 (47.6) & 67.9 (46.5) & $-3.7$ \\
mbfa-latin, $\beta = 15$ (default) & 71.2 (50.0) & 67.5 (47.0) & 70.0 (49.5) & 70.3 (49.3) & 69.0 (48.2) & $-3.7$ \\
mbfa-latin, $\beta = 30$ & 70.5 (49.2) & 67.5 (46.9) & 69.6 (49.0) & 69.8 (48.7) & 68.7 (47.9) & $\mathbf{-3.0}$ \\
mbfa-latin-noEnglish, $\beta = 15$ & 70.9 (49.8) & 67.5 (47.0) & 69.9 (49.5) & 70.2 (49.2) & 68.8 (48.2) & $-3.4$ \\
Charsiu W2V2-FS & 75.8 (44.0) & 69.8 (39.6) & 71.9 (42.3) & 74.7 (43.2) & 71.4 (40.8) & $-6.0$ \\
MAPS\textsuperscript{\textdaggerdbl} & 74.8 (47.9) & 68.4 (41.7) & 71.8 (46.9) & 70.5 (43.4) & 68.1 (41.7) & $-6.7$ \\
MFA english\_us\_arpa & \textbf{83.0} (58.9) & \textbf{75.5} (52.1) & \textbf{79.4} (56.5) & \textbf{81.9} (57.9) & \textbf{75.8} (52.2) & $-7.6$ \\
MFA english\_mfa & 82.6 (\textbf{60.2}) & 75.3 (\textbf{53.6}) & 78.8 (\textbf{57.0}) & 81.2 (\textbf{58.9}) & 75.5 (\textbf{53.1}) & $-7.2$ \\
\bottomrule
\end{tabular*}
\end{table}

\begin{table*}[t]
\caption{\Fone{25} (\%) under transcript errors, averaged over the error kinds at each rate (Clean: 0\%), and its slope: the least-squares fit of \Fone{25} against the error rate, in points per 10\% of phones or words corrupted ($R^2 \ge 0.96$ for every row). Best per corpus and column in bold (for slopes, the shallowest) on TIMIT and Buckeye, not counting MAPS. \textsuperscript{\textdaggerdbl}Scored on its held-out clips (1,498 TIMIT test clips; 3,144 Buckeye segments of four speakers). WildPhonemes English: 95 clips that no system, MAPS included, was trained on; its reference was made by correcting BFA's alignment, so it favours BFA, mbfa-latin above all (Section~\ref{sec:res-wild}), and none of its columns is bolded. Every system, onset weight and error kind is in Supplementary Section~S3.}\label{tab:errslope}
\footnotesize
\begin{tabular*}{\textwidth}{@{\extracolsep{\fill}}lrrrrrrrr@{}}
\toprule
 & & \multicolumn{4}{c}{Phone errors} & \multicolumn{3}{c}{Word errors} \\
\cmidrule(lr){3-6} \cmidrule(l){7-9}
System & Clean & 5\% & 10\% & 20\% & Slope & 10\% & 20\% & Slope \\
\midrule
\multicolumn{9}{@{}l}{\emph{TIMIT}} \\
CUPE-3-en, $\beta = 15$ (default) & 77.6 & 76.3 & 74.9 & 72.5 & $-2.54$ & 74.0 & 70.5 & $-3.55$ \\
CUPE-3-en, $\beta = 30$ & 77.6 & 76.5 & 75.4 & \textbf{73.5} & $-2.06$ & 74.3 & \textbf{71.0} & $-3.29$ \\
CUPE-3-en, $\beta = 0$ & 75.5 & 73.7 & 71.9 & 68.8 & $-3.35$ & 71.7 & 68.0 & $-3.74$ \\
CUPE-3-mls7, $\beta = 15$ & 72.1 & 71.2 & 70.3 & 68.9 & $-1.62$ & 69.7 & 66.8 & $-2.66$ \\
mbfa-latin, $\beta = 15$ (default) & 72.5 & 71.5 & 70.8 & 69.5 & $-1.43$ & 70.1 & 67.5 & $-2.48$ \\
mbfa-latin, $\beta = 30$ & 73.5 & 72.8 & 72.3 & 71.3 & $\mathbf{-1.10}$ & 71.3 & 69.0 & $-2.28$ \\
mbfa-latin-noEnglish, $\beta = 15$ & 72.7 & 72.0 & 71.2 & 70.0 & $-1.34$ & 70.5 & 68.2 & $\mathbf{-2.27}$ \\
Charsiu W2V2-FS & 76.0 & 74.5 & 72.9 & 70.2 & $-2.90$ & 71.2 & 66.7 & $-4.65$ \\
MAPS\textsuperscript{\textdaggerdbl} & 80.4 & 77.7 & 74.9 & 69.3 & $-5.55$ & 74.9 & 69.7 & $-5.35$ \\
MFA english\_us\_arpa & \textbf{80.8} & \textbf{78.3} & \textbf{75.8} & 71.2 & $-4.79$ & \textbf{75.3} & 70.2 & $-5.27$ \\
MFA english\_mfa & 79.1 & 76.9 & 74.6 & 70.4 & $-4.38$ & 74.1 & 69.4 & $-4.84$ \\
\midrule
\multicolumn{9}{@{}l}{\emph{Buckeye}} \\
CUPE-3-en, $\beta = 15$ (default) & 76.5 & 75.1 & 73.7 & 71.2 & $-2.64$ & 72.3 & 68.4 & $-4.04$ \\
CUPE-3-en, $\beta = 30$ & 75.1 & 73.9 & 72.7 & 70.7 & $-2.17$ & 71.3 & 67.9 & $-3.59$ \\
CUPE-3-en, $\beta = 0$ & 77.7 & 75.8 & 74.0 & 70.7 & $-3.50$ & 72.9 & 68.4 & $-4.64$ \\
CUPE-3-mls7, $\beta = 15$ & 70.3 & 69.3 & 68.3 & 66.5 & $-1.89$ & 67.3 & 64.6 & $-2.88$ \\
mbfa-latin, $\beta = 15$ (default) & 71.2 & 70.2 & 69.2 & 67.5 & $-1.87$ & 68.1 & 65.2 & $-3.01$ \\
mbfa-latin, $\beta = 30$ & 70.5 & 69.7 & 68.9 & 67.6 & $\mathbf{-1.43}$ & 67.8 & 65.2 & $\mathbf{-2.60}$ \\
mbfa-latin-noEnglish, $\beta = 15$ & 70.9 & 70.0 & 69.1 & 67.6 & $-1.64$ & 68.0 & 65.3 & $-2.79$ \\
Charsiu W2V2-FS & 75.8 & 73.9 & 72.0 & 68.3 & $-3.75$ & 69.8 & 64.3 & $-5.77$ \\
MAPS\textsuperscript{\textdaggerdbl} & 74.8 & 72.2 & 69.7 & 64.8 & $-4.99$ & 69.1 & 63.8 & $-5.49$ \\
MFA english\_us\_arpa & \textbf{83.1} & \textbf{80.6} & \textbf{78.2} & \textbf{73.6} & $-4.72$ & \textbf{77.1} & \textbf{71.8} & $-5.61$ \\
MFA english\_mfa & 82.6 & 80.1 & 77.7 & 73.0 & $-4.78$ & 76.7 & 71.6 & $-5.50$ \\
\midrule
\multicolumn{9}{@{}l}{\emph{WildPhonemes English (reference seeded by BFA)}} \\
CUPE-3-en, $\beta = 15$ (default) & 73.5 & 72.1 & 71.2 & 69.3 & $-2.05$ & 70.2 & 67.7 & $-2.89$ \\
CUPE-3-en, $\beta = 30$ & 74.8 & 73.5 & 72.6 & 70.8 & $-1.93$ & 71.5 & 69.3 & $-2.72$ \\
CUPE-3-en, $\beta = 0$ & 70.2 & 68.6 & 67.0 & 64.8 & $-2.65$ & 66.7 & 64.6 & $-2.79$ \\
CUPE-3-mls7, $\beta = 15$ & 69.6 & 68.4 & 67.4 & 66.2 & $-1.66$ & 66.9 & 64.8 & $-2.38$ \\
mbfa-latin, $\beta = 15$ (default) & 82.3 & 80.4 & 78.4 & 74.1 & $-4.14$ & 77.9 & 74.0 & $-4.17$ \\
mbfa-latin, $\beta = 30$ & 81.2 & 79.4 & 78.0 & 74.6 & $-3.26$ & 77.3 & 73.8 & $-3.71$ \\
mbfa-latin-noEnglish, $\beta = 15$ & 73.7 & 73.0 & 71.5 & 69.1 & $-2.36$ & 70.8 & 68.5 & $-2.58$ \\
Charsiu W2V2-FS & 64.1 & 63.3 & 62.4 & 60.1 & $-2.01$ & 60.8 & 58.2 & $-2.96$ \\
MAPS & 54.0 & 51.6 & 51.2 & 46.6 & $-3.53$ & 50.5 & 47.5 & $-3.23$ \\
MFA english\_us\_arpa & 68.5 & 66.5 & 64.5 & 60.8 & $-3.81$ & 64.3 & 61.2 & $-3.63$ \\
MFA english\_mfa & 67.9 & 66.1 & 64.3 & 60.8 & $-3.58$ & 63.7 & 61.1 & $-3.40$ \\
\bottomrule
\end{tabular*}
\end{table*}

\paragraph{Accuracy falls slowly.} Averaged over the error kinds, every system's F1 falls almost linearly with the error rate (Fig.~\ref{fig:errors}), on TIMIT and Buckeye alike (Table~\ref{tab:errslope}); Table~\ref{tab:errors} breaks the drop down by error kind, and every rate is in Supplementary Section~S3.

\paragraph{The onset term contributes.} The onset term's evidence comes from the audio, not from the transcript. Without it ($\beta = 0$), CUPE-3-en and mbfa-latin degrade more on both corpora. A larger weight ($\beta = 30$) gives both models a shallower slope, by 0.3--0.5 points per 10\% of phones corrupted. On TIMIT that also makes it the most accurate setting under errors. On Buckeye, where the larger weight costs 0.7--1.4 points on the clean transcript, the smaller losses do not make up that difference even at 20\% errors, except for mbfa-latin, where the two settings end level; and $\beta = 0$, the more accurate setting for CUPE-3-en on clean Buckeye transcripts (Supplementary Section~S8), falls below $\beta = 15$ at 20\% phone errors.

\paragraph{The robustness does not come from the target language.} mbfa-latin-noEnglish, which never heard English, degrades least of the four models at their defaults, under both phone and word errors and on both corpora. CUPE-3-en, the same network as CUPE-3-mls7 continued on English, degrades somewhat more but is still the more accurate of the two at 20\% phone errors, so English training raised its accuracy more than it lowered its robustness.

\paragraph{The same picture on WildPhonemes.} On the English clips of WildPhonemes, crowdsourced read speech recorded outside the studio, CUPE-3-en, CUPE-3-mls7 and mbfa-latin-noEnglish degrade at rates similar to those on the hand-labelled corpora (Table~\ref{tab:errslope}). The exception is mbfa-latin, whose clean score is the highest because the reference favours it most, and whose slope is much steeper than on TIMIT and Buckeye, plausibly because an error that moves it away from its clean alignment also costs it agreement with its own seed.

\paragraph{Confidence partly flags corrupted transcripts.} At a 10\% error rate on TIMIT, mean per-phone confidence separates corrupted from clean clips at an AUROC of 0.67--0.78 (CUPE-3-en) for inserted and substituted phones and words, but not for deletions (about 0.5). On Buckeye's conversational segments it is weaker, 0.57--0.65, and again about 0.5 for deletions. A deleted phone leaves its audio to neighbours that are really there, so nothing in the alignment looks unsupported, whereas an inserted or substituted phone is placed on audio that does not match it, and its posterior says so. For single onsets on clean transcripts confidence is a weaker signal (AUROC 0.56--0.63 for an onset that is unmatched or more than 50\,ms off). It is a coarse flag for text that the audio does not support, not a calibrated error estimate for each phone.

\subsection{Robustness to audio corruption}\label{sec:res-audio}

Section~\ref{sec:res-errors} changed the transcript and kept the audio. Here we change the audio and keep the transcript, in two ways: a sound that is not in the transcript inserted into a pause, and a word's audio replaced by a different word. Every condition is still scored against the hand labels.

\paragraph{Protocol.} We use the full corpora of Section~\ref{sec:main}: all 6,300 TIMIT clips and all 255 Buckeye interviews. Every system hears the same edited audio. Both edits join audio from two recordings, and where the joins lie matters. A join is an abrupt spectral change, which is what the onset term rewards, so joins placed exactly on a word's edges would make the test partly measure the joins (Supplementary Section~S4). Every join here lies in quiet audio instead: at least 15\,ms inside a pause or a stop closure, and so at least 15\,ms from any reference boundary, where the audio within 10\,ms is at least 25\,dB below the active speech level in both recordings (the mean power of the clip's 20\,ms frames within 35\,dB of its loudest), with a 10\,ms raised-cosine cross-fade. For the \emph{untranscribed sound}, one pause per segment (an internal one where the segment has one) receives 250--800\,ms of another recording of the same speaker played backwards (median 531\,ms) or, on Buckeye, one of the speaker's own filled pauses (``uh'', ``um''). Everything after it moves by exactly its length; every reference boundary is kept, and every predicted onset inside the sound is a false alarm. For the \emph{word swap}, a word whose two edges can both be joined in quiet audio is replaced by another word of the same speaker with a different spelling and 0.8--1.25 times its duration. This is possible for only 1,568 of TIMIT's 54,387 words (2.9\%) and 1,347 of Buckeye's 284,809 (0.5\%), nearly all next to a pause or a stop closure, and all of them are used. The swapped word's own reference boundaries, and predicted onsets more than 20\,ms inside it, are dropped, in the clean audio too; its two edges are scored. Each edit has a control that makes the same kind of joins without changing the content: room tone of the same length shaped like the pause's own audio, which only lengthens the pause, and the word kept with the middles of the quiet stretches beside it replaced. We run BFA's four models at the default $\beta = 15$ and, as in Section~\ref{sec:res-errors}, CUPE-3-en at $\beta = 30$ and $0$ and mbfa-latin at $\beta = 30$; MFA \texttt{english\_us\_arpa} with one folder per speaker; and Charsiu W2V2-FS. MAPS is scored on the clips it did not see, TIMIT's test set and the four held-out Buckeye speakers, on which every other system was also run (Supplementary Section~S4).

\paragraph{Reading Tables~\ref{tab:audio} and~\ref{tab:swap}.} Every number except the hit rates in Table~\ref{tab:swap} is a change in percentage points against the same clips with clean audio, so $0$ means no effect and more negative is worse. In Table~\ref{tab:audio}, $\Delta$\Fone{25} is the change of \Fone{25} over all boundaries of the corpus with reversed speech inserted. \emph{Inside} is the share of insertion sites at which the system places at least one onset more than 20\,ms inside the inserted sound, i.e.\ assigns some phone of the transcript to audio that contains none (lower is better). The last four columns ask where the damage falls: the change in the share of reference boundaries hit within 25\,ms, for the last boundary before the sound, the first and second after it, and those five or more boundaries after it. In Table~\ref{tab:swap}, the swapped word's two edges are hit within 25\,ms at the rate in \emph{Clean} with the original audio and in \emph{Swapped} with the other word's audio; $\Delta$ is the difference. \emph{Control} is the same difference when the joins are made but the word is kept, so it measures the joins alone, and \emph{Next} is the change for the first boundary after the word.

\begin{table}[t]
\caption{Reversed speech inserted into one pause per segment, in points against the clean audio (Section~\ref{sec:res-audio}). $\Delta$\Fone{25}: over all boundaries. Inside: share (\%) of inserted sounds in which the system starts at least one phone of the transcript (more than 20\,ms from either edge of the sound), although the sound contains none; lower is better. Before, 1st, 2nd, 5th+: change in the share of reference boundaries hit at 25\,ms, by position relative to the sound. Best per corpus and column in bold, not counting MAPS. \textsuperscript{\textdaggerdbl}MAPS is scored on its held-out clips only (1,680 TIMIT test clips; four Buckeye speakers), and its per-position changes were not computed. Every system, fillers, room tone, positions and the held-out clips are in Supplementary Section~S4.}\label{tab:audio}
\footnotesize
\begin{tabular*}{\tblwidth}{@{\extracolsep{\fill}}lrrrrrr@{}}
\toprule
 & & & \multicolumn{4}{c}{$\Delta$ hit@25 by position} \\
\cmidrule(l){4-7}
System & $\Delta$\Fone{25} & Inside & Before & 1st & 2nd & 5th+ \\
\midrule
\multicolumn{7}{@{}l}{\emph{TIMIT (6,300 sites)}} \\
CUPE-3-en, $\beta = 15$ (default) & $-2.0$ & 44.3 & $-8.3$ & $-12.4$ & $\mathbf{-16.4}$ & $-0.5$ \\
CUPE-3-en, $\beta = 30$ & $-2.3$ & 49.1 & $-10.1$ & $-15.8$ & $-19.2$ & $-0.6$ \\
CUPE-3-en, $\beta = 0$ & $\mathbf{-1.7}$ & \textbf{42.7} & $-7.7$ & $\mathbf{-6.1}$ & $-17.9$ & $\mathbf{-0.3}$ \\
CUPE-3-mls7, $\beta = 15$ & $-3.0$ & 53.5 & $-11.6$ & $-22.4$ & $-21.1$ & $-1.1$ \\
mbfa-latin, $\beta = 15$ (default) & $-6.1$ & 87.1 & $-22.3$ & $-30.0$ & $-37.6$ & $-2.6$ \\
mbfa-latin, $\beta = 30$ & $-6.3$ & 90.0 & $-20.7$ & $-29.0$ & $-33.8$ & $-2.9$ \\
mbfa-latin-noEnglish, $\beta = 15$ & $-6.6$ & 89.6 & $-19.4$ & $-32.3$ & $-34.6$ & $-2.9$ \\
Charsiu W2V2-FS & $-7.3$ & 73.3 & $-25.2$ & $-36.7$ & $-43.8$ & $-3.7$ \\
MAPS\textsuperscript{\textdaggerdbl} & $-2.6$ & 51.5 & -- & -- & -- & -- \\
MFA english\_us\_arpa & $-8.6$ & 55.1 & $\mathbf{-1.1}$ & $-17.8$ & $-35.4$ & $-13.9$ \\
\midrule
\multicolumn{7}{@{}l}{\emph{Buckeye (7,911 sites)}} \\
CUPE-3-en, $\beta = 15$ (default) & $-1.2$ & 73.4 & $-11.7$ & $\mathbf{-15.3}$ & $\mathbf{-15.4}$ & $-1.6$ \\
CUPE-3-en, $\beta = 30$ & $-1.5$ & 78.5 & $-12.2$ & $\mathbf{-15.3}$ & $-15.5$ & $-2.2$ \\
CUPE-3-en, $\beta = 0$ & $\mathbf{-1.1}$ & 69.3 & $-11.6$ & $-15.7$ & $-16.5$ & $\mathbf{-1.3}$ \\
CUPE-3-mls7, $\beta = 15$ & $-1.7$ & 85.1 & $-13.3$ & $-18.1$ & $-16.4$ & $-2.5$ \\
mbfa-latin, $\beta = 15$ (default) & $-2.1$ & 93.6 & $-12.5$ & $-19.2$ & $-21.3$ & $-3.6$ \\
mbfa-latin, $\beta = 30$ & $-2.3$ & 95.7 & $-11.1$ & $-16.6$ & $-19.0$ & $-4.0$ \\
mbfa-latin-noEnglish, $\beta = 15$ & $-2.3$ & 95.0 & $-10.8$ & $-20.2$ & $-20.5$ & $-4.1$ \\
Charsiu W2V2-FS & $-1.7$ & \textbf{66.8} & $-14.1$ & $-22.3$ & $-21.2$ & $-1.9$ \\
MAPS\textsuperscript{\textdaggerdbl} & $-0.6$ & 68.8 & -- & -- & -- & -- \\
MFA english\_us\_arpa & $-3.0$ & 73.9 & $\mathbf{-3.7}$ & $-24.6$ & $-34.0$ & $-9.2$ \\
\bottomrule
\end{tabular*}
\end{table}

\begin{table}[t]
\caption{A word's audio replaced by another word of the same speaker, transcript unchanged (Section~\ref{sec:res-audio}). Share (\%) of the word's two edges hit at 25\,ms with the clean and the swapped audio, and its change in points; Control: the same change when the joins are made but the word is kept; Next: change for the first boundary after the word. Best per corpus and column in bold, not counting MAPS or the controls. \textsuperscript{\textdaggerdbl}MAPS is scored on its held-out clips only (428 TIMIT words; 138 Buckeye words); it hits the edges of a Buckeye word between two pauses about 5\% of the time even in the clean audio (Section~\ref{sec:res-audio}), so its Buckeye swap is not compared. Every system and condition is in Supplementary Section~S4.}\label{tab:swap}
\footnotesize
\begin{tabular*}{\tblwidth}{@{\extracolsep{\fill}}lrrrrr@{}}
\toprule
 & \multicolumn{3}{c}{Word edges hit@25 (\%)} & & \\
\cmidrule(lr){2-4}
System & Clean & Swapped & $\Delta$ & Control & Next \\
\midrule
\multicolumn{6}{@{}l}{\emph{TIMIT (1,568 words)}} \\
CUPE-3-en, $\beta = 15$ (default) & 87.6 & 76.3 & $-11.3$ & $+0.5$ & $-3.8$ \\
CUPE-3-en, $\beta = 30$ & 87.6 & 79.5 & $-8.1$ & $+0.4$ & $-4.2$ \\
CUPE-3-en, $\beta = 0$ & 84.0 & 67.7 & $-16.3$ & $+0.4$ & $-4.3$ \\
CUPE-3-mls7, $\beta = 15$ & 82.9 & 72.8 & $-10.1$ & $-0.2$ & $-3.3$ \\
mbfa-latin, $\beta = 15$ (default) & 90.7 & 82.3 & $-8.4$ & $-0.4$ & $-3.2$ \\
mbfa-latin, $\beta = 30$ & \textbf{90.8} & \textbf{85.8} & $-5.0$ & $+0.1$ & $-2.0$ \\
mbfa-latin-noEnglish, $\beta = 15$ & 89.3 & 81.4 & $-7.9$ & $-0.4$ & $-3.7$ \\
Charsiu W2V2-FS & 48.4 & 47.3 & $\mathbf{-1.1}$ & $-0.7$ & $-3.1$ \\
MAPS\textsuperscript{\textdaggerdbl} & 84.3 & 52.2 & $-32.1$ & $-1.3$ & -- \\
MFA english\_us\_arpa & 73.0 & 67.8 & $-5.2$ & $+0.7$ & $\mathbf{+5.6}$ \\
\midrule
\multicolumn{6}{@{}l}{\emph{Buckeye (1,347 words)}} \\
CUPE-3-en, $\beta = 15$ (default) & 72.1 & 59.0 & $-13.1$ & $0.0$ & $-0.8$ \\
CUPE-3-en, $\beta = 30$ & 72.8 & 59.9 & $-12.9$ & $-0.5$ & $-0.8$ \\
CUPE-3-en, $\beta = 0$ & 68.1 & 48.6 & $-19.5$ & $+0.5$ & $-1.4$ \\
CUPE-3-mls7, $\beta = 15$ & 66.5 & 53.8 & $-12.7$ & $0.0$ & $-1.7$ \\
mbfa-latin, $\beta = 15$ (default) & 79.4 & 76.6 & $-2.8$ & $+0.4$ & $-0.4$ \\
mbfa-latin, $\beta = 30$ & 79.1 & 76.4 & $\mathbf{-2.7}$ & $+1.1$ & $\mathbf{+0.7}$ \\
mbfa-latin-noEnglish, $\beta = 15$ & \textbf{80.9} & \textbf{77.3} & $-3.6$ & $+1.1$ & $-1.4$ \\
Charsiu W2V2-FS & 57.5 & 47.4 & $-10.1$ & $+0.4$ & $-5.2$ \\
MAPS\textsuperscript{\textdaggerdbl} & 4.4 & 10.9 & -- & -- & -- \\
MFA english\_us\_arpa & 71.0 & 55.6 & $-15.4$ & $+0.2$ & $-3.1$ \\
\bottomrule
\end{tabular*}
\end{table}

\paragraph{An untranscribed sound.} Every system places onsets inside the inserted sound at many of the sites (where ideally it shouldn't), the std models at the most (Table~\ref{tab:audio}): with no blank class, a phone of the transcript is often the best available explanation of speech-like audio. Room tone of the same length, which only lengthens the pause, costs every system little, and a real filled pause costs slightly less than reversed speech (Supplementary Section~S4), so it is the sound's content that does the damage. The systems differ in where the damage goes. MFA hands the sound mostly to the following words, and boundaries five or more positions later are still affected, so one sound disturbs the rest of the segment. BFA shares the sound between the words on either side, and little of it reaches boundaries five or more positions away. The onset term is the one part of BFA that this test penalises: a larger $\beta$ places more onsets inside the sound, whose own spectral changes look like phone onsets.

\paragraph{A word that does not match its audio.} Swapped words are few, so F1 over all boundaries barely moves, and Table~\ref{tab:swap} scores the swapped words' own edges. Every system hits them less often after the swap, while the controls barely change them, so the loss comes from audio that does not match the transcript, not from the joins. Without the onset term BFA loses more, as under transcript errors. Beyond the edges the effect is local: the next boundary changes by a few points, and boundaries four or more away barely at all (Supplementary Section~S4). Charsiu's clean edges are hit far less often than other systems', which limits what its small loss says. MAPS adds silence only at the two ends of a transcript, so it places the onset after an internal pause badly, and nearly every swapped Buckeye word lies between two pauses.

\subsection{From CTC to hat targets}\label{sec:res-trail}

BFA was developed from CUPE-2i, the CTC model of Table~\ref{tab:main}, through a sequence of runs, each scored by phone-onset F1 on TIMIT (Table~\ref{tab:trail}). The table is a short summary of a longer trail: each row stands for a stage that was explored over many runs, and most stages change more than one thing at once, so ablation runs are also included for single-variable experiments. The code of every stage is public in the git history of the repository (Data availability). Read in order, the rows show where the gain over CTC decoding comes from.

\paragraph{The CTC output loses the encoder's timing.} Under CTC with a blank class, a spectral-stream branch and more training data (stage~B) raised \Fone{25} from about 50 to about 67, after which further changes to the CTC model gained little. A probe of the layers of one such model shows where the timing is lost (B1--B4). Onsets peak-picked from its convolutional spectral-stream layer (B2) are about as accurate as the best CTC configurations (B, B1), whereas onsets peak-picked after the transformer (B3), or read from the CTC output (B4), are about 20 points worse at 25\,ms. The convolutional layers localise the boundaries, and the layers above them, whose objective constrains only which phones are emitted and in what order, do not preserve that localisation.

\paragraph{Hat targets and the segmental search.} Replacing the transformer and the blank class with a CNN trained on hat targets (Section~\ref{sec:hat}) raises \Fone{25} by about 19 points over a CTC run with the same 100\,h of data and the same number of epochs, both with a plain decoder (B5 to C0). The plain decoder is the CTC decoder of BFA~1.x \citep{rehman2026bfa1x}, with its optional onset snapping off: a frame-level Viterbi search on the raw log-posteriors, with blank states between the phones, in which a phone may take a single frame. For the hat model, whose blank class is never a training target, the blank states can only cost it, so C0 is if anything a low estimate. This step changes the objective and the architecture at once, but the two can be separated: a CTC model with the same CNN-only trunk, data and decoder (B6) gains only about 2 points over B5 and still trails C0 by about 17, so this step's gain comes from the objective rather than the trunk. B5, B6 and C0 share the 13.125\,ms frame rate. The segmental search of Section~\ref{sec:topology} replaces the plain decoder with phones of at least $D$ frames under a soft duration prior, scaled-likelihood emissions (Section~\ref{sec:emissions}) and optional silence states, and adds about 11 points to the same network without retraining (C2), so a good part of the gain lies in the decoder; the onset term is added only at C4. Training on the search's own alignments (C3), more data and the onset term (C4), and decoding at the final defaults (C) bring the network to the released CUPE-3-en-base. Decoding the C4 network without the onset term costs about 2 points of \Fone{25} and about 11 of \Fone{10} (C4a). No single step accounts for the final accuracy: it comes from the objective and the segmental search together, with the onset term adding most at 10\,ms.

\paragraph{Frame rate.} Moving the hat model from the 13.125\,ms to the 5\,ms grid (C0 to C1) leaves \Fone{25} unchanged and gains about 2 points of \Fone{10}. In the other direction, rounding the released model's onsets to a coarser grid (C6a--C6d) costs under one point of \Fone{25}, even at 20\,ms, but several points of \Fone{10}, about three at MFA's 10\,ms grid. The 5\,ms grid therefore matters at the tightest tolerance (Section~\ref{sec:main}), and the gap at 25\,ms between the CTC rows of Table~\ref{tab:main} and BFA is not a matter of frame rate.

\paragraph{Context.} A further single-variable ablation is not in the table. Adding an attention layer, which widens the receptive field from 39 to 79\,ms, does not help either (\Fone{25} of 75.5 with it, 75.9 without): wider context does not make the boundaries more precise, which is why the trunk omits CUPE's transformer layers (Section~\ref{sec:encoder}).

\begin{table}[t]
\caption{From CUPE-2i to CUPE-3: phone-onset F1 (\%) on all 6,300 TIMIT clips at 10, 25 and 50\,ms (Section~\ref{sec:res-trail}). Every run is trained on LibriSpeech. Each row summarises a stage of the development; every stage is in the repository's git history. Rows B1--B4 probe the layers of one CTC model; rows C6a--C6d round the onsets of row C to a coarser grid. Rows A, C and C5 are the CUPE-2i with CTC, CUPE-3-en-base and CUPE-3-en rows of Table~\ref{tab:main}.}\label{tab:trail}
\footnotesize
\begin{tabular*}{\tblwidth}{@{\extracolsep{\fill}}lllrrrr@{}}
\toprule
Stage & Change & Run & Frame (ms) & \Fone{10} & \Fone{25} & \Fone{50} \\
\midrule
A   & CTC with blank, CNN + transformer          & CUPE-2i        & 13.125 & 22.8 & 50.1 & 77.3 \\
B   & + spectral-stream branch, 360\,h           & CTC-Spec-360h  & 13.125 & 38.6 & 67.2 & 83.0 \\
B1  & CTC onsets snapped to the spectral stream  & CTC-Spec-Full  & 13.125 & 46.4 & 69.2 & 80.4 \\
B2  & Peak-picked from the spectral-stream layer & CTC-Spec-Full  & 13.125 & 49.4 & 66.9 & 76.8 \\
B3  & Peak-picked after the transformer          & CTC-Spec-Full  & 13.125 & 21.4 & 46.1 & 69.4 \\
B4  & CTC output                                 & CTC-Spec-Full  & 13.125 & 21.9 & 47.9 & 73.0 \\
B5  & CTC, 100\,h, 20 epochs (matched to C0)     & CTC-Base-100h  & 13.125 & 20.1 & 42.5 & 65.4 \\
B6  & CTC, with blank, CNN only; plain Viterbi   & CTC-CNN-100h   & 13.125    & 19.6 & 44.7 & 68.6 \\
\midrule
C0  & Hat targets, no blank, CNN only; plain Viterbi & Hat-CNN-100h & 13.125         & 30.2 & 61.5 & 78.9 \\
C1  & + 5\,ms frame grid                            & Hat-CNN-100h & 5         & 32.6 & 61.4 & 78.7 \\
C2  & + segmental search, no retraining          & Hat-CNN-100h   & 5      & 54.8 & 72.2 & 84.1 \\
C3  & + trained on segmental alignments          & Hat-CNN-Seg    & 5      & 56.4 & 73.2 & 84.3 \\
C4  & + 500\,h, onset term                       & Hat-CNN-500h   & 5      & 58.7 & 76.0 & 85.4 \\
C4a & C4 decoded without the onset term          & Hat-CNN-500h   & 5      & 47.3 & 74.0 & 85.9 \\
C   & C4 as released, decoder defaults           & CUPE-3-en-base & 5      & 59.5 & \textbf{78.6} & 88.2 \\
C5  & Occupancy targets, from CUPE-3-mls7        & CUPE-3-en      & 5      & \textbf{59.9} & 78.2 & 87.9 \\
\midrule
C6a  & C rounded to a 10\,ms grid                 & CUPE-3-en-base & 10     & 56.8 & 78.4 & 88.2 \\
C6b  & C rounded to a 13.125\,ms grid             & CUPE-3-en-base & 13.125 & 56.3 & 78.3 & 88.2 \\
C6c  & C rounded to a 16\,ms grid                 & CUPE-3-en-base & 16     & 55.1 & 78.2 & 88.2 \\
C6d  & C rounded to a 20\,ms grid                 & CUPE-3-en-base & 20     & 51.9 & 77.8 & 88.1 \\
\bottomrule
\end{tabular*}
\end{table}

\subsection{Errors by manner class}\label{sec:manner}

\begin{table}[t]
\caption{Mean absolute error (ms) of matched onsets on TIMIT, by the manner class of the phone that starts at the boundary. Best per column in bold. Last row: each class's share (\%) of mbfa-latin's training phones (FLEURS, Latin-script group, silence and noise excluded); affricates (1.1\%) and glottals (1.0\%) make up the rest; per language in Supplementary Section~S5.}\label{tab:manner}
\small
\begin{tabular*}{\tblwidth}{@{\extracolsep{\fill}}lrrrrrrrrr@{}}
\toprule
 & Vowel & Vl.\ stop & Vd.\ stop & Vl.\ fric. & Vd.\ fric. & Nasal & Lateral & Rhotic & Glide \\
\midrule
CUPE-3-en & 12.2 & 11.8 & 17.2 & \textbf{10.9} & 14.0 & 13.0 & 20.4 & \textbf{17.5} & 16.8 \\
mbfa-latin & 17.0 & 14.4 & 17.7 & 11.5 & 14.3 & 12.9 & 19.9 & 19.0 & \textbf{14.1} \\
MFA english\_us\_arpa & \textbf{11.3} & \textbf{11.1} & \textbf{14.5} & 13.4 & \textbf{12.7} & \textbf{10.8} & \textbf{17.7} & 18.7 & 16.7 \\
\midrule
\emph{mbfa-latin training share} & \emph{43.4} & \emph{12.5} & \emph{6.7} & \emph{8.0} & \emph{3.8} & \emph{10.5} & \emph{4.7} & \emph{5.3} & \emph{3.1} \\
\bottomrule
\end{tabular*}
\end{table}

Laterals and rhotics are the hardest classes for every system (Table~\ref{tab:manner}): their onsets are gradual formant transitions with no acoustic edge, so where the boundary lies is partly a labelling convention. CUPE-3-en is most precise on voiceless fricatives, whose onset is an abrupt rise of high-frequency noise, exactly what the onset term rewards, and less precise on voiced stops, voiced fricatives and nasals, across whose onsets voicing usually continues. mbfa-latin's vowel onsets are the least precise of the three systems', plausibly because its vowel classes must cover the vowel systems of 36 languages: vowels are its most frequent training class, but spread over 74 distinct tokens. Conversely, glides are among its rarest training classes, yet it places their onsets most precisely of the three systems. How often a class occurs in training thus says little about how well its onsets are placed, which matters for languages whose phones are rare in their group's training data.

\subsection{Cost}\label{sec:res-cost}

On one GPU, BFA aligns both corpora at a real-time factor of 0.01 or less. Without a GPU, the network dominates, and its cost follows the padded audio rather than the speech, so padding segments to 10\,s rather than 30\,s makes it two to three times faster. Timings of BFA and MFA on one machine, with every configuration and per-stage split, are in Supplementary Section~S6.

\section{Discussion}\label{sec:discussion}

\subsection{What a new language needs}\label{sec:newlang}

To align a language, BFA needs a G2P model that covers it and a script group whose tokens cover its phones. It needs no pronunciation dictionary and no boundary labels, and the models that never heard English (Section~\ref{sec:res-unseen}) suggest that it need not have heard the language at all, as long as it has heard its phones in other languages. This is what a short receptive field buys: a network that sees 39\,ms of audio models how phones sound, not which words or phone sequences a language uses, so what it learns from one language carries over to another that shares its phones. The 20 languages that can be phonemised into a group's tokens without having been in its training data (Section~\ref{sec:inventory}) would be aligned in the same way, but we have not measured them. English is a well-resourced and phonologically typical member of the Latin-script group; whether the transfer holds as well for a language whose phones are rare in its group's training data, or whose phonology differs more from the rest of its group, is untested. Studies of cross-language HMM alignment suggest that it will not hold uniformly: transfer is good between close languages and worse between distant ones \citep{young2023forced,dicanio2013}, tone and the size of the inventory affect which source model works best \citep{wu2024cross}, and the most accurate model on its own language need not be the most accurate on another \citep{leinonen2022semiautomatic}. \citet{chodroff2025comparing} found that a large and varied English model matched language-specific models trained on small amounts of target-language data, and called for tests of truly multilingual models. A script-group model trained on dozens of languages is such a model, and we speculate that the breadth of its training data, as much as its short receptive field, is what lets it serve languages it has not heard; this paper tests that only on English. The manner-class results (Section~\ref{sec:manner}) suggest that rarity alone matters less than whether a class has a clear acoustic edge. In practice the limiting component is the front end: a language without a usable G2P model, or whose orthography the G2P handles poorly, cannot be aligned well whatever the acoustic model knows (Section~\ref{sec:limitations}).

\subsection{Reference boundaries and the aligner bias}\label{sec:aligner_bias}

Labelling phone boundaries from scratch is slow, so corpora with phone-level references usually start from an automatic alignment and have annotators correct it. An annotator moves a boundary that looks wrong and keeps one that looks acceptable, so wherever the seed was acceptable the reference keeps the seeding aligner's placement. Correction typically moves only a minority of onsets by more than 20\,ms \citep{agmon2025challenges}, and aligner output differs from annotators in systematic ways, for example in starting words later and leaving short pauses between them \citep{leinonen2022semiautomatic}; a seeded reference keeps such tendencies wherever the annotator let them stand. An aligner that shares that aligner's model family, phone set or dictionary then inherits part of its score, whatever the audio says. This leaves a dilemma for every evaluation of aligners, ours included: the references we can score against are not neutral between the systems evaluated, and how much of a score comes from the audio and how much from a likeness to the seed cannot be told from the scores alone.

The two corpora differ in how strongly this applies. TIMIT's boundaries were first placed automatically and then corrected by hand \citep{zue1990timit}, by an acoustic-phonetic aligner rather than an HMM, and with conventions of their own, such as separate closure and release segments for stops, that neither BFA nor MFA reproduces. Buckeye's audio was first aligned by an HMM aligner with TIMIT-trained phone models \citep{pitt2007buckeye}. Transcribers chose the best-fitting pronunciation from that aligner's ARPAbet variant dictionary and moved boundaries they judged misplaced; a boundary they found acceptable kept its HMM position. That can favour an HMM aligner in three ways unrelated to the audio: shared placement conventions, a shared phone set, and a benchmark that HMM aligners are developed against \citep{mfamodels}. BFA's training uses no boundary from any aligner or annotator, so whatever a reference inherited from its seed, BFA has not learned it. We therefore report both corpora in full and draw no conclusion from the difference between them. A reference that is neutral between aligners would have to be annotated from a seed that favours none of them. WildPhonemes is not one: BFA seeded its phone boundaries and, with MFA, its word edges, which is why we use it only as a check (Section~\ref{sec:res-wild}).

Evaluating aligners whose phone sets differ from each other and from the reference forces a choice, and it is unavoidable for an aligner meant for many languages. Label-aware matching, which pairs boundaries only between corresponding phones, needs a mapping between phone sets and penalises systems whose inventories split or merge phones differently from the reference. Label-free matching, as used here, tests only timing, but could in principle reward a system that places many onsets. One-to-one matching limits this, since extra onsets cannot stack on one boundary and each unmatched one lowers precision, and the predicted-to-reference ratio, which we report for every system, shows how many onsets each places. The same ratio exposes the two-token affricates and diphthongs of the std set (Section~\ref{sec:inventory}). BFA's segmentation is contiguous (Section~\ref{sec:topology}), so onset and offset boundaries coincide and there are no gaps between phones to resolve.

\subsection{One onset weight, three regimes}\label{sec:onsetweight}

The onset term is evidence from the audio alone, and its value depends on how far the other evidence can be trusted. On read speech with a correct transcript, it sharpens boundaries that the posteriors already place approximately. When the transcript is wrong, or the audio of a word does not match it, it anchors the surviving boundaries to the audio. Where the audio holds a sound that is not in the transcript, it pulls onsets into that sound (Section~\ref{sec:res-audio}). No single weight is best in all three regimes, and the default, $\beta = 15$ for every model, is a compromise between them (Supplementary Section~S8). Where transcripts of read speech are unreliable, $\beta = 30$ is the safer choice (on conversational speech its smaller losses do not make up for its lower accuracy on clean transcripts, Section~\ref{sec:res-errors}). In training, the one matched pair of runs we have favours keeping the term on whatever the decoding weight (Supplementary Section~S8): without it the self-alignment has no evidence beyond its own posteriors with which to correct its targets.

\subsection{Limitations}\label{sec:limitations}

\begin{itemize}
\item Boundary accuracy is verified impartially only on English. In the other languages it is human-verified but partial to BFA: WildPhonemes (Section~\ref{sec:res-wild}) serves only as a check, because BFA seeded its reference (every inner phone boundary and, with MFA, every word edge), it is not an established benchmark, and it is small, with read prompts, 24--98 clips per language and one to three languages per script group. Although it serves its purpose as a verification on in-the-wild data, it does not evaluate alignment accuracy properly. The field is short of publicly available datasets that could be used to evaluate multilingual alignment claims \citep[see also][]{deng2025speech}. A few datasets exist behind paid licences (as TIMIT does), but given the biases and convention mismatches that persist in Buckeye and TIMIT, we took the route of our own annotation and verification, as carefully as we could. In future, we plan to expand at least three of the non-English subsets to one hour each.
\item G2P errors propagate, a G2P model gives the standard pronunciation of a word whether or not the speaker used it \citep{agmon2025challenges}, and digits, symbols and abbreviations are not normalised in the std path.
\item Every phone of the transcript is placed. Words absent from the audio are squeezed in and only flagged through confidence, as the definition of forced alignment requires.
\item The front end also labels each phone with tone, stress and length, taken from the G2P, and the std models were trained with an auxiliary tone head (22 Chao contours, loss weight 0.3, masked for languages without tone). None of this is evaluated here: the tone labels are lexical, not acoustic, and the tone head is not read at inference. Acoustic tone labelling is left to future work.
\end{itemize}

\section{Conclusion}\label{sec:conclusion}

Our goal was to see if neural forced alignment could produce phone boundaries in a variety of languages without the need for boundary labels or a pronunciation dictionary. This is possible because of two design decisions. First, the boundary is incorporated into the posteriors: hat-shaped targets cause neighboring phones to intersect at their common boundary, and the targets are constructed using the same segmental search employed during inference, allowing the model to bootstrap solely from transcripts, similar to an HMM aligner. Substituting CTC's sharp posteriors and blank class with this objective, alongside the segmental search and its onset term, resulted in BFA boundaries being precise enough for phonetic applications. Second, the acoustic model processes merely a few tens of milliseconds of audio each frame, which is insufficient to grasp words or phonotactics, so it focuses on modeling the sounds of phones. By using a G2P front end instead of a dictionary and having one inventory for each writing system, a single model caters to every language within a script group.

The models that never heard English are the most direct evidence for the second point. They align English nearly as well as models trained with it, so what an acoustic model learns about a phone in one language carries over to another language that shares it. Training data in the target language still improve accuracy, but they are not required. The script-group models, adapted from the Latin-script model on 7--40\,h of speech each, passed a check in fourteen languages against human-verified boundaries, although that reference was seeded in part by BFA and is partial to it. Under strain the aligner degrades gradually: errors in the transcript or audio that the transcript does not describe disturb the boundaries near the error and leave the rest of the segment largely intact. The acoustic-change onset term illustrates the trade-offs of using audio evidence alone. It anchors boundaries when the transcript cannot be trusted, but it also pulls onsets into sounds that the transcript does not describe. No single weight suits every regime, so the weight should be chosen to fit the material.

The evaluation has its own limits. Every phone-level reference available to us was seeded by an automatic aligner, and a reference keeps its seed wherever annotators found it acceptable. Scores against such references are therefore partly a measure of likeness to the seed, and a fair evaluation of multilingual aligners will need references annotated from neutral seeds, in more languages and in greater depth than any public resource now offers. Beyond the acoustic model, the limiting component for a new language is the front end. Where no usable G2P model exists, or where the G2P handles the orthography poorly, the alignment can be no better than the phones it is given. Expanding the non-English references, assessing languages whose phones are rare in their group's training data, and labelling tone from the acoustics are the next steps. We release the aligner, the models, the evaluation harness and the development history, so that this work can be reproduced and extended without retraining from scratch.

\printcredits

\section*{Data availability}
The aligner (\texttt{bournemouth-forced-aligner}; version 2.1.0; \citealp{rehman2026bfa21}), the evaluation harness and the configuration of every reported run are available at \url{https://github.com/tabahi/bournemouth-forced-aligner}, whose git history holds the code of every development stage of Section~\ref{sec:res-trail}; all 14 models are on the Hugging Face Hub (\url{https://huggingface.co/Tabahi}). The BFA library is released under the GPL-3.0 license, and the models under the AGPL-3.0 license. TIMIT is distributed by the Linguistic Data Consortium (LDC93S1) and the Buckeye corpus by the Ohio State University; neither can be redistributed. FLEURS, LibriSpeech, MLS and Common Voice are publicly available. WildPhonemes, with its audio, transcripts and word and phone reference alignments in Praat TextGrid and JSON, is available on Zenodo under the CC0-1.0 license (\url{https://doi.org/10.5281/zenodo.24484757}) \citep{ar2026_wildphonemes}.

\section*{Acknowledgements}
This work was supported by Bournemouth University. The funder had no role in the study design; in the collection, analysis and interpretation of data; in the writing of the article; or in the decision to submit it for publication.

\bibliographystyle{cas-model2-names}
\bibliography{bfa_paper}

\end{document}